\documentclass[3p,times,twocolumn]{elsarticle}

\usepackage{lineno,hyperref}
\modulolinenumbers[5]

\journal{Journal of Ad Hoc Networks}

\usepackage{graphicx}
\usepackage{mathptmx}
\usepackage{tabularx}
\usepackage{ragged2e}
\usepackage[singlelinecheck=false]{caption}
\usepackage[T1]{fontenc}
\usepackage[numbers]{natbib}
\usepackage{amsmath}
\usepackage{amssymb}
\usepackage{textcomp}
\usepackage{booktabs}

\usepackage{pdfpages}
\usepackage{graphicx,import}

\begin{document}

\begin{frontmatter}

\title{T-ROME: A Simple and Energy Efficient Tree Routing Protocol for Low-Power Wake-up Receivers}

\author[affil1]{T. Kumberg\corref{cor1}}
\cortext[cor1]{Corresponding author. Tel.: +49 761 203 7243.}
\ead{timo.kumberg@imtek.de}
\author[affil1]{M. Schink}
\author[affil1]{L. M. Reindl}
\author[affil2]{C. Schindelhauer}

\address[affil1]{University of Freiburg -- IMTEK -- Laboratory for Electrical Instrumentation, Georges-Koehler-Allee 106. 79110 Freiburg}

\address[affil2]{University of Freiburg -- IIF -- Computer Networks and Telematics, Georges-Koehler-Allee 51, 79110 Freiburg, Germany}

\begin{abstract}
Wireless sensor networks are deployed in many monitoring applications but still suffer from short lifetimes originating from limited energy sources and storages. Due to their low-power consumption and their on-demand communication ability, wake-up receivers represent an energy efficient and simple enhancement to wireless sensor nodes and wireless sensor network protocols. In this context, wake-up receivers have the ability to increase the network lifetime. In this article, we present T-ROME, a simple and energy efficient cross-layer routing protocol for wireless sensor nodes containing wake-up receivers. The protocol makes use of the different transmission ranges of wake-up and main radios in order to save energy by skipping nodes during data transfer. With respect to energy consumption and latency, T-ROME outperforms existing protocols in many scenarios. Here, we describe and analyze the cross layer multi-hop protocol by means of a Markov chain model that we verify using a laboratory test setup.
\end{abstract}


\end{frontmatter}


\section{Introduction}

Wireless sensor networks are used in many applications like environmental monitoring, home automation, smart manufacturing, infrastructure monitoring and many others. In this context, a wireless sensor network usually consists of many small self-powered sensor nodes that measure their environment, process data and communicate it to other nodes or to a base station \cite{puccinelli2005}. Message transmission can be done via single-hop transmissions or via multi-hop communication resulting in complex network topologies. 

The most critical parameter of a wireless sensor node is its energy requirement \cite{heba2013} which is vastly dominated by the power required for communication. A lot of research was already done on efficient MAC protocols to reduce power consumption and collisions and to increase the throughput of a wireless network \cite{pei2013}. The authors of \cite{pei2013} categorize MAC protocols into four groups: asynchronous, synchronous, frame-slotted, and multi-channel protocols. Asynchronous and synchronous protocols are based on duty-cycling, where nodes switch between sleep and active states in order to save energy. To establish a communication link in synchronous protocols like S-MAC or T-MAC, each participating node has to be awake at the same time. This necessitates clock synchronization messages. Asynchronous protocols like B-MAC or WiseMAC nodes use preamble sampling in combination with duty-cycling to detect the beginning of a communication. To minimize collisions frame-slotted protocols allocate different time slots to nearby nodes. Multi-channel protocols use cross-channel communication to realize higher throughput. All these MAC protocols have in common that their energy requirement is linked to the duration of their sleep periods. Longer sleep periods result in lower energy consumption but also in communication latencies. In addition, these MAC protocols require a certain amount of overhead to organize themselves \cite{heba2013}. 

Recently, wireless sensor networks \cite{hoflinger2014, ban2016, kumbergPLSE, gamm2012smart, kumberg2016wake} have been upgraded with low-power wake-up receivers. These wake-up receivers have marginal power consumption and wake up the sensor node if a dedicated signal has been received. So, low-power wake-up receivers can greatly reduce the power consumption of wireless sensor nodes, by eliminating the idle listening time and at the same time reduce communication delays to achieve an almost latency free communication \cite{blanck2015}.

According to \cite{heba2013}, wake-up radios can be categorized into two groups, active and passive wake-up receivers. Passive wake-up receivers harvest their wake-up energy directly from the wake-up message itself, whereas active wake-up receivers require a permanent, yet very low, power supply. In this approach, a wireless sensor node usually incorporates two radio receivers, the main radio for data communication and a second one for receiving wake-up messages \cite{heba2013}. A sensor node wakes up only when it receives a wake-up message and then it turns on its communication radio.

Another advantage of wireless sensor nodes with wake-up receivers is their enhanced robustness. Clock synchronization is obsolete and nodes may be reset at any time, for example, if a fatal software error occurred. Existing networks can be easily enhanced by new nodes, even if the network is running on low duty cycle periods \cite{kumbergPLSE}. Furthermore, extracting data from the network can be done without much delay, as messages are transmitted almost instantly. 

Although \cite{ban2015} speak of a paradigm shift for wireless sensor protocols with integrated wake-up transceivers, there exist two major challenges \cite{heba2013, ban2015}: First, active wake-up receivers show a higher sensitivity compared to passive ones \cite{heba2013}, but their sensitivity is still lower compared to that of state-of-the-art main communication radio transceivers. Secondly, sending wake-up messages may cost more energy than sending of communication messages. Table~\ref{tab:typ_transmitter} shows the typical sensitivity of some commonly used radio transmitters and their current consumption during transmit state. In Table~\ref{tab:wakeup_receiver} sensitivity and power consumption of some state-of-the-art wake-up receivers are shown. The discrepancy between main radio and wake-up receiver sensitivity is clearly obvious as is the power consumption.

\begin{table}[h!]
  \caption{Receiver (RX) Sensitivity at 868 MHz and transmit (TX) currents at +10 dBm for some typical RF transmitters.}
\label{tab:typ_transmitter}
  \begin{tabularx}{\hsize}{p{1.8cm} X X X }
\toprule
    RF Transceiver & RX Sensitivity [dBm]& TX current [mA]\\
\midrule
    Si4468 & -104 & 19.7 \\
    CC1200 & -107 & 36 \\
    CC1101 & -95 & 30  \\
    SPIRIT1 & -105 & 21 \\
\bottomrule
  \end{tabularx}
\end{table}

\begin{table}[h!]
  \caption{Non-exhaustive list of wake-up receivers, their sensitivity and power consumption.}
\label{tab:wakeup_receiver}
  \begin{tabularx}{\hsize}{p{4cm} X X X }
\toprule
    Wake-up receiver & Sensitivity [dBm]& Power [\textmu W]\\
\midrule
   Magno and Benini \cite{ma2014} & -55 & 1.3 \\
  Nilsson and Svensson \cite{nil2013} & -47 & 2.3  \\
   Gamm et al. \cite{gamm2012} & -52 & 5.6 \\
   Hambeck et al. \cite{hamb2011} & -71 & 2.4 \\
    
\bottomrule
  \end{tabularx}
\end{table}

Here, we present a cross-layer multi-hop wake-up routing protocol that combines wake-up and communication radios. The wireless sensor nodes are based on the works of \cite{gamm2012, kumberg1}. Due to the smaller transmission range of wake-up receivers compared to that of the main radio, data and wake-up transmissions are realized by a multi-hop routing protocol that supports sending wake-up messages and data. The protocol stack consists of several layers. The lowest layer is responsible for the waking up of neighboring nodes. The second layer handles single-hop message transmissions and the top layer routes messages and forwards wake-up signals along multiple hops.

The presented work in this paper is organized as follows. In Section~\ref{sec:related_work} we review existing network protocols that support the use of wake-up receivers. In~Section \ref{sec:sensor_node} we take a look at current wake-up receiver designs and present the wireless sensor node that is used in this research. In~Section \ref{sec:network_protocol} we introduce the proposed multi-hop wake-up routing protocol in detail and analyze its current consumption as well as the occurrence of false wake-ups in Section \ref{sec:exp_ana}. In Section \ref{sec:analysis}, we introduce Markov models of the proposed algorithm as well as for CTP-WUR and a naive communication algorithm. The models are verified and performance and energy requirements of the aforementioned protocols are compared and analyzed in Section \ref{sec:results}. Finally, outlook and conclusions can be found in~Section \ref{sec:outlook}.

\section{Related Work}
\label{sec:related_work}

\subsection{Wake-up Transceiver}
\label{sec:sensor_node}

\begin{figure}[ht]
\centering
		\includegraphics[scale=0.4]{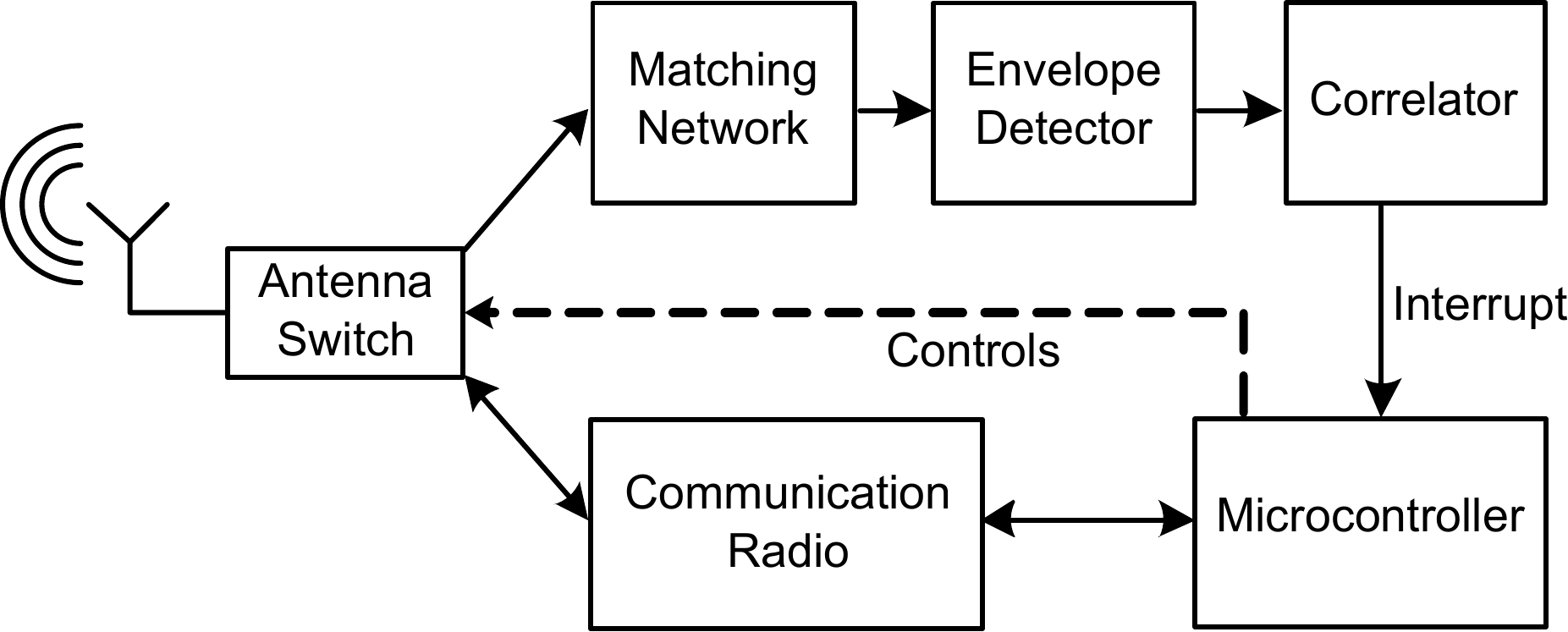}
		\caption{Schematic of wireless sensor node including a wake-up receiver.}
		\label{fig:wakeup_schematic}
\end{figure}

\begin{figure}[ht]
\centering
		\includegraphics[scale=0.5]{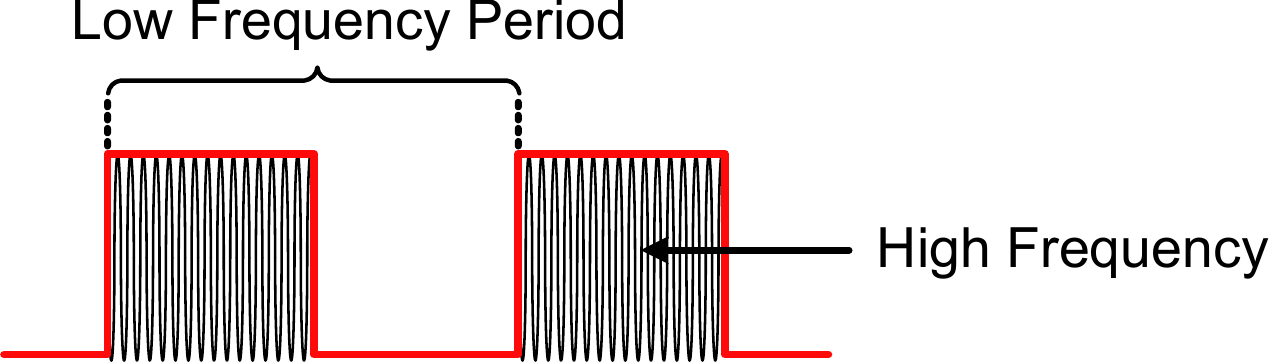}
		\caption{The low-frequency wake-up message (red) is modulated on the high-frequency carrier signal by On Off Keying.}
		\label{fig:modulation}
\end{figure}

Generally, a low-power wake-up receiver consists of an envelope detector and a correlator as sketched in Figure \ref{fig:wakeup_schematic} that shows schematically a wireless sensor node including a wake-up receiver. The envelope detector demodulates the high-frequency (HF) carrier signal to achieve a low-frequency (LF) wake-up signal as sketched in Figure \ref{fig:modulation} \cite{gamm2012} that depicts an On-Off-Keying modulated wake-up signal. The correlator analyzes the LF signal, to verify the validity of a wake-up message. In that case, the main microcontroller of the sensor node is woken up by an interrupt and, depending on the embedded software, a sensor reading might be initiated or the antenna is connected to the main radio to establish further communications. A matching network might be necessary to match the impedances of antenna and wake-up receiver. 

Blanck et al. presented \cite{blanck2015} an overview of current low-power transceivers. In respect to energy consumption the range goes from highly integrated concepts that require 0.1 \textmu W \cite{roberts201298nw, oh2013116nw} to several solutions between 10 and 1000 \textmu W \cite{blanck2015}. Only a few receivers are in the range of 1 to 10 \textmu W. Common to all receivers in the range below of 10 \textmu W is that they use On-Off-Keying modulated wake-up messages. This is due to the simplified and energy efficient hardware design that can be used in this particular case. For example, the envelope detector is merely composed of diodes and capacitors and as a correlator, a comparator is used \cite{ma2014, mosh2015, ansari2009, spenza2015, marinkovic2011nano}. 


\subsection{Protocols}
\label{sec:protocols}

Although wake-up receivers have many advantages and writer frequently reported devices in wireless sensor networks \cite{hoflinger2014, ban2016, kumbergPLSE, gamm2012smart}, there do not exist many MAC or routing protocols that support their use and the majority of existing protocols are only limited to simulations. Some existing protocols for wake-up receivers support single-hop communication only, like E2RMAC \cite{jain2007}, WUR-MAC \cite{mahl2009}, RTWAC \cite{ansari2009} and GWR-MAC \cite{karvonen2014}. These protocols show superior energy requirements compared to synchronous or asynchronous MAC protocols but their performance is only based on simulation results. The main feature of E2RMAC and WUR-MAC protocols is to use the wake-up signal as an RTS/CTS mechanism to avoid the hidden terminal problem. In RTWAC all nodes have a unique and a common wake-up address to support broadcasting and dedicated messages. But the purpose of wake-up messages is only to trigger an event, for example, a sensor reading, at the receiver node. Data communication is realized by a more common CSMA/CA MAC protocol that is not further specified, using the main radio. 

The protocols presented in \cite{marinkovic2011} and \cite{gamm2012smart} were tested in real applications but are also limited to single-hop communications. Similar to those protocols but designed for body area networks is the work of \cite{mosh2015}. The protocol introduces additionally a random back-off time to avoid collisions. The protocol as presented in \cite{marinkovic2011} combines wake-up messages and a low duty cycle TDMA based MAC protocol \cite{marinkovic2009} to increase flexibility. Performance evaluation is done by comparing the proposed protocol with and without a wake-up radio. 

Recently \cite{petri2014} presented a novel wake-up receiver design together with two flooding protocols FLOOD-WUP and GREEN-WUP. FLOOD-WUP uses different broadcast addresses to forward messages to receivers that are not in range of the first transmitter and to avoid the reception of multiple messages. GREEN-WUP includes additional information about harvested energy at a node coded in its address and nodes with higher energy levels are preferred relay nodes. Evaluation of both protocols is only performed on the basis of simulation and the authors do not evaluate the power requirements of the proposed protocols. \cite{sutton2015} presented ZIPPY, an on-demand multi-hop flooding technique based on wake-up receivers. ZIPPY is extensively tested in a laboratory testbed and shows latencies in the range of tens of milliseconds to broadcast multi-hop messages. 

CTP-WUR, a cross-layer routing protocol for wake-up receivers presented in \cite{basagni2016ctp} introduces relaying of wake-up messages by using flagged wake-up messages to inform the receiver about the intended multi-hop wake-up. The relaying node forwards the wake-up call to its parent node that itself starts to wait for data from the first node. In case the node woke up due to a false wake-up, the node goes back to sleep after a predefined time has passed and no data is received. The protocol allows for relaying of one wake-up message at maximum. Wake-up messages are not acknowledged but  successful data transfer is indicated by an acknowledgment from receiver to the sender. Data communication is done via the CTP routing protocol \cite{gnawali2013ctp}. The authors of \cite{chen2015mh} present MH-REACH-Mote, a node based on the Tmote-Sky platform in combination with a wake-up receiver. In their scenario, communication is done from a mobile sink to fix nodes. Wake-up messages are relayed from the nearest fix node to the ones further away from the mobile sink. The protocol assumes no collisions and an existing communication link from the fix sources to the mobile sink. 

The protocol presented in \cite{blanck2012} uses low-power wake-up receivers to create clusters of sensor nodes that exhibit similar sensor readings and only cluster heads transmit information to the sink. Sensor readings and cluster configuration messages are encoded into wake-up messages. The protocol shows promising results for applications with many similar sensor readings. The authors of \cite{spenza2015} introduce ALBA-WUR a cross-layer network protocol that supports the use of wake-up receivers. Wake-up addresses are chosen dynamically from a set of predefined wake-up addresses, depending on packet size and on historical node performance. In simulations, ALBA-WUR showed superior power consumption and latency as compared to ALBA-R a geographic cross-layer routing protocol with contention-based MAC \cite{petrioli2014alba}.

Of course, the quality of wireless links can change quickly due to changes in the environment \cite{srinivasan2008beta}. To achieve a robust, reliable and efficient routing, state-of-the-art wireless network protocols like CTP \cite{gnawali2013ctp} estimate the current link quality between nodes and adjust their routing paths accordingly. The link quality estimation can either be achieved by incorporating information from different network layers like the number of received acknowledgments and the link quality indicator provided by the radio or it can be based on the $\beta$-factor \cite{srinivasan2008beta} that measures the burstiness of a wireless link. While link estimation is a common technique in traditional wireless network protocols, it is not standard in all used wireless routing protocols that are based on wake-up receivers since an accurate and timely link quality estimation requires a certain amount of control messages (beacons) to be sent. This is energy-wise expensive due to the high costs of wake-up messages. 

ALBA-WUR, for example, calculates the link quality by taking into account how many packets have been lost on a specific link in the past. This achieves a good average link quality information but cannot resemble fast or short link quality changes. To avoid collisions and to improve the reliability WUR-MAC chooses dynamically one out of several available channels of the 2.4 GHz ISM band for wake-up transmissions. To choose a channel, the protocol keeps track of all channels used in neighboring nodes for communication and then takes randomly one of the remaining channels for its own communication. This approach does not avoid collisions and like ALBA-WUR only calculates an average channel usage without the possibility to react on rapid channel fluctuations. In T-ROME we introduce a parameter to assist the sender in order to dynamically choose the best next hop node based on multiple values like distance to the source and link quality estimation. This also enables route adjusting on rapidly changing link conditions.

With the aim to reduce the number of transmissions from the source to sink, and as such to increase network performance, opportunistic routing protocols rely on broadcasting data packets to several nodes (the set of candidates) to forward a message from a source to sink \cite{boukerche2015opportunistic}. Usually, the most appropriate forwarder is chosen out of the set of candidates based on local and end-to-end metrics. Local metrics are based on link conditions and geographic positions of the sensor nodes, while end-to-end metrics are usually based on link properties between source and destination \cite{boukerche2015opportunistic}. 

In traditional opportunistic routing, it is necessary that each node of the candidate set receives the broadcast data packet and answers back to the sender. The authors of \cite{boukerche2015opportunistic} categorize this candidate coordination into two groups, either being based on control messages or on time-coded sending of data packets. In the latter, a node's priority is proportional to a time period it waits until it forwards a data packet. If a node overhears a data transmission from another node, it knows that it does not have the highest priority and does not forward the packet. The drawback of this method is that multiple data packets may be transmitted in case a node does not hear the transmission of another one. In case the candidate coordination is based on control messages, acknowledgments or the RTS-CTS frame can be used \cite{boukerche2015opportunistic}. In both approaches, the time to sent an acknowledgment or the CTS message is proportional to a node's priority and if a node overhears an acknowledgment or a CTS message from another node, it backs off. The difference between the two approaches is that in acknowledgment based candidate coordination, the data packet is received by all possible candidates and in the RTS-CTS approach, the data is sent to the most appropriate candidate only. 

FLOOD-WUP realizes opportunistic routing according to the acknowledgment based approach but forwarding is done after a random period of time has passed. To avoid multiple transmissions of the same data packet, each node changes its wake-up address upon reception of a data packet. Although changing of the wake-up address follows a certain sequence, it can happen that a node loses the proper sequence and additional control packets are required \cite{petri2014}. The opportunistic routing in GREEN-WUP is similar to that of FLOOD-WUP but wake-up addresses are additionally based on the current energy level of a sensor node and the source node goes to sleep after it sent the initial wake-up sequence. A possible relay node has to wake up the source by using a unicast wake-up packet that was initially provided by the source. Due to this, GREEN-WUP requires additional wake-up packets that are usually expensive with respect to energy.  

The cross-layer routing protocol presented in this work is based on existing nodes, in contrast to most of the protocols above. We realized the RTS/CTS messages similar to those presented above but additionally our protocol supports multi-hop communication and forwarding mechanisms similar to those presented in ALBA-WUR and GREEN-WUP but other than the latter protocols, our protocol does not use flooding. The protocols presented in \cite{basagni2016ctp} and \cite{chen2015mh} use relaying of wake-up messages similar to our proposed solution but use only one relay node, whereas the number of relay nodes in T-ROME is not limited. Additionally, in T-ROME we implemented the possibility to include a set of decision parameter that can be used to dynamically optimize the relaying process and to choose the optimal relay node similar to the opportunistic RTS-CTS approach shown above, but the candidate set is established during the routing itself. Furthermore, T-ROME introduces a mechanism to send several data packets in a row along an existing link.

\section{Wireless Sensor Node}
\label{sec:sensor_node}

\begin{figure}[ht]
\centering
		\includegraphics[scale=0.3]{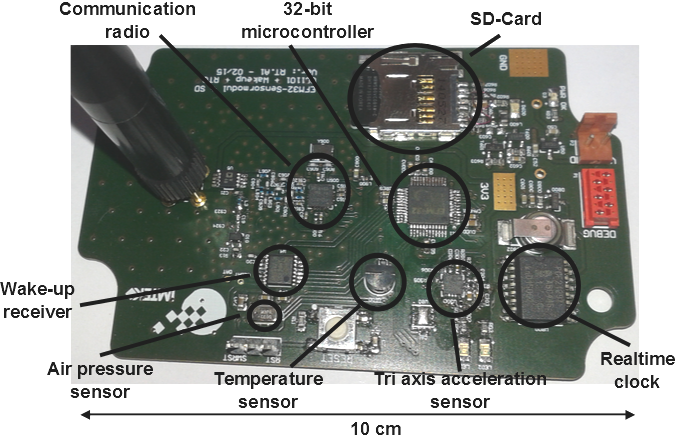}
		\caption{Photo of sensor node with wake-up receiver.}
		\label{fig:node}
\end{figure}

\begin{figure*}[ht]
\centering
		\includegraphics[scale=0.5]{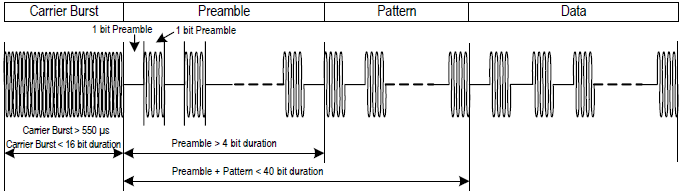}
		\caption{Wake-up pattern of the AS3932 LF wake-up receiver. Pattern and data are optional.}
		\label{fig:as_pattern}
\end{figure*}

The wireless nodes used in this work are based on the sensor node introduced in \cite{gamm2012, kumberg1}. Figure \ref{fig:node} shows a photo of the implemented node. The microcontroller utilized on the boards is a 32 bit EFM32G222F128 manufactured by SiliconLabs running at 14 MHz. It provides several low power states to reduce energy consumption. In run mode it draws around 2.5 mA and 0.9 \textmu A in Deep Sleep Mode. Including all peripherals, the sensor node requires around 4.0 mA in run mode. The communication radio is a CC1101 from Texas Instruments. It has a current consumption of 34.2 mA when transmitting at +12 dBm output power at 868 MHz and around 16.4 mA when transmitting at 0 dBm. Its sensitivity is approximately in between -95 to -104 dBm, depending on the data rate. The 125 kHz LF receiver (AS3932) from austriamicrosystems has a current consumption of around 3 \textmu A in listening mode. It correlates the incoming signal to a pre-configured address and creates an interrupt if send and stored addresses match. In combination with matching network and an envelope detector, the wake-up receiver has a sensitivity around -51 dBm \cite{gamm2012, kumberg1}. Figure \ref{fig:as_pattern} shows the wake-up pattern required to wake-up the AS3932 chip consisting of carrier burst, preamble and optional address and pattern. The CC1101 transceiver generates the wake-up pattern by modulating the pattern on the 868 MHz signal by means of an On-Off-Keying modulation as presented in Section \ref{sec:sensor_node}.

In addition to some common sensors, the node is equipped with a high precision realtime clock (PCF2129T) and a MicroSD card that can be switch off by the microcontroller. A monopole antenna with a gain of approximately 1.5 dBi is used.

Friis transmission equation can be used to calculate the freespace transmission distances for wake-up and main radio. But knowing that the wireless transmission range is additionally affected by multi-path propagation effects like reflection, scattering, and diffraction a more accurate model that includes multi-path fading can be used to estimate the range, as given in \cite{kvaksrud2008range}. Using the Matlab function provided there, the communication range of the main radio can be estimated to be well above 300 m when sending at + 0 dBm output power. Transmitting at +12 dBm, the wake-up range can be calculated likewise to be around 45 m.  Experiments conducted in \cite{kumberg1} using a similar sensor node compared to the one introduced here, implies the wake-up range to be around 45 m even for sending at 10 dBm output power.

\subsection{False Positive and False Negative Wake-ups}
\label{sec:false_wakeup}

Due to their low power consumption and the fact that they listen always on incoming signals, wake-up receivers are prone to false positive and false negative wake-ups.  In the case of false positive wake-ups, a receiver detects a valid signal although the wake-up message was not dedicated to it. False negative wakeups occur when a wake-up receiver stays asleep although a wake-up message was sent to it. Both kinds of false wake-ups can result from interferences on the wireless channel and can possibly lead to an increased power consumption and communication delays.

Experimentally, the occurrence of false negative wake-ups can be measured for example by counting how many valid wake-ups a receiver detected out of the number of sent valid wake-up messages. The false positive wake-up rate can be experimentally measured by counting how often a wake-up receiver detects a valid wake-up message although the message does not contain a valid address. 

To reduce the occurrence of false positive and false negative wake-ups some wake-up receivers use active or passive input filter \cite{petri2014}, which includes a correlator unit that analyses the received wake-up messages and only creates a wake-up signal in case the addresses match \cite{petri2014, oh2013116nw, ma2014}, or make use of manchester or similarly encoded wake-up signals \cite{gamm2012, kumberg1, ansari2009}. 

\section{Network Protocol}
\label{sec:network_protocol}

As already introduced in Tables \ref{tab:typ_transmitter} and \ref{tab:wakeup_receiver},  the sensitivity of wake-up receivers is lower than that of communication radios. This means that data can be sent over longer distances than wake-up messages as shown in Section \ref{sec:sensor_node}. 

\begin{figure}[ht]
\centering
		\includegraphics[scale=0.6]{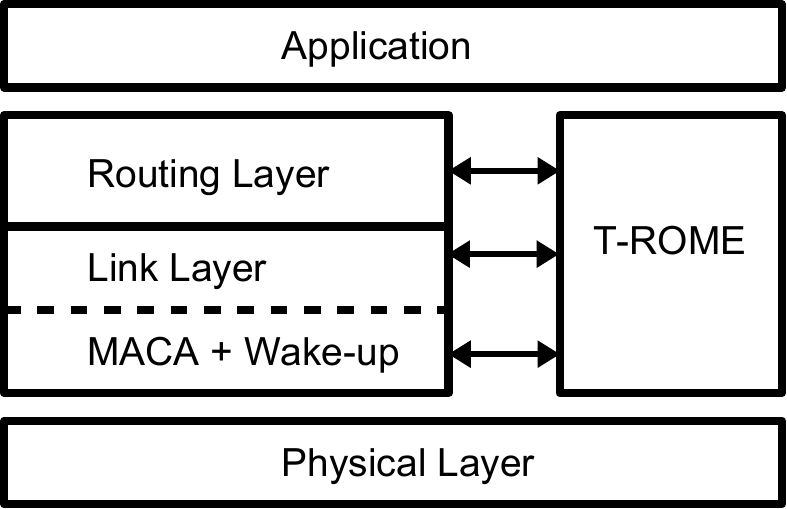}
		\caption{Communication layer stack consisting of physical layer, link layer, routing layer and application. The cross-layer protocol T-ROME supports functions in the link and the routing layer as depicted in the figure. The Wake-up is embedded in the link layer and supports the RTS/CTS scheme based on MACA to reduce packet collisions.}
		\label{fig:stack}
\end{figure}

Due to this, T-ROME is a cross-layer protocol, as visualized in Figure \ref{fig:stack}. Above the physical layer is the link layer that supports single-hop transmissions and waking up of neighboring nodes. This is realized basically by using an RTS/CTS message exchange to reduce packet collisions as introduced in MACA \cite{karn1990maca}. In this context, a wake-up message works also as an RTS and the wake-up acknowledgment as the CTS command. The routing layer routes messages along multiple hops according to a static routing table implemented on each node. Following sections introduce the cross-layer protocol and corresponding data packets in more details. The application runs above the communication layers. 

\subsection{T-ROME Protocol}

\begin{figure}[ht]
\centering
		\includegraphics[scale=0.7]{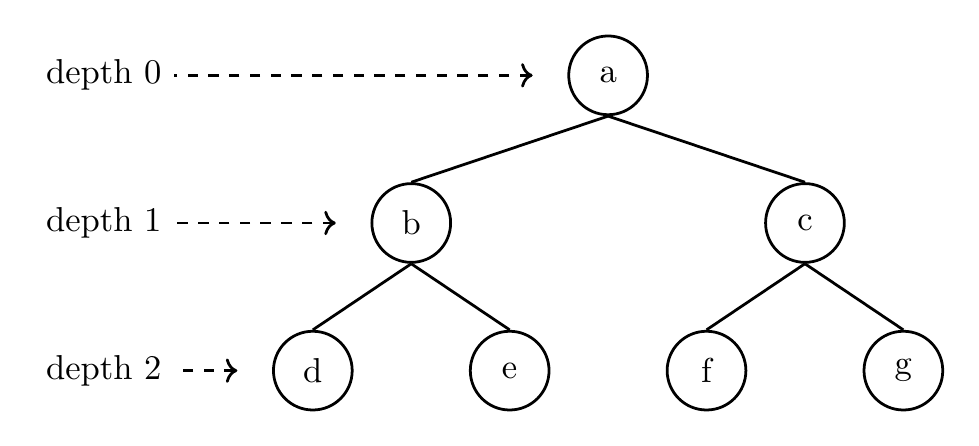}
		\caption{Schematic of a simple tree routing protocol with nodes a to g. Communication is only possible from child to parent for example from \textit{node b} to \textit{node a}.}
		\label{fig:tree_top}
\end{figure}

The static cross-layer protocol is based on the simple Tree Routing algorithm \cite{qiu2009}. In this protocol messages can be passed only from child to parent nodes as depicted in Figure \ref{fig:tree_top}. Every node of a certain depth $i$ is able to communicate with a node of depth $i-1$ and vice versa. For example \textit{node b} is able to communicate to \textit{node a}. 

\begin{figure*}[ht]
\centering
		\includegraphics[scale = 0.8]{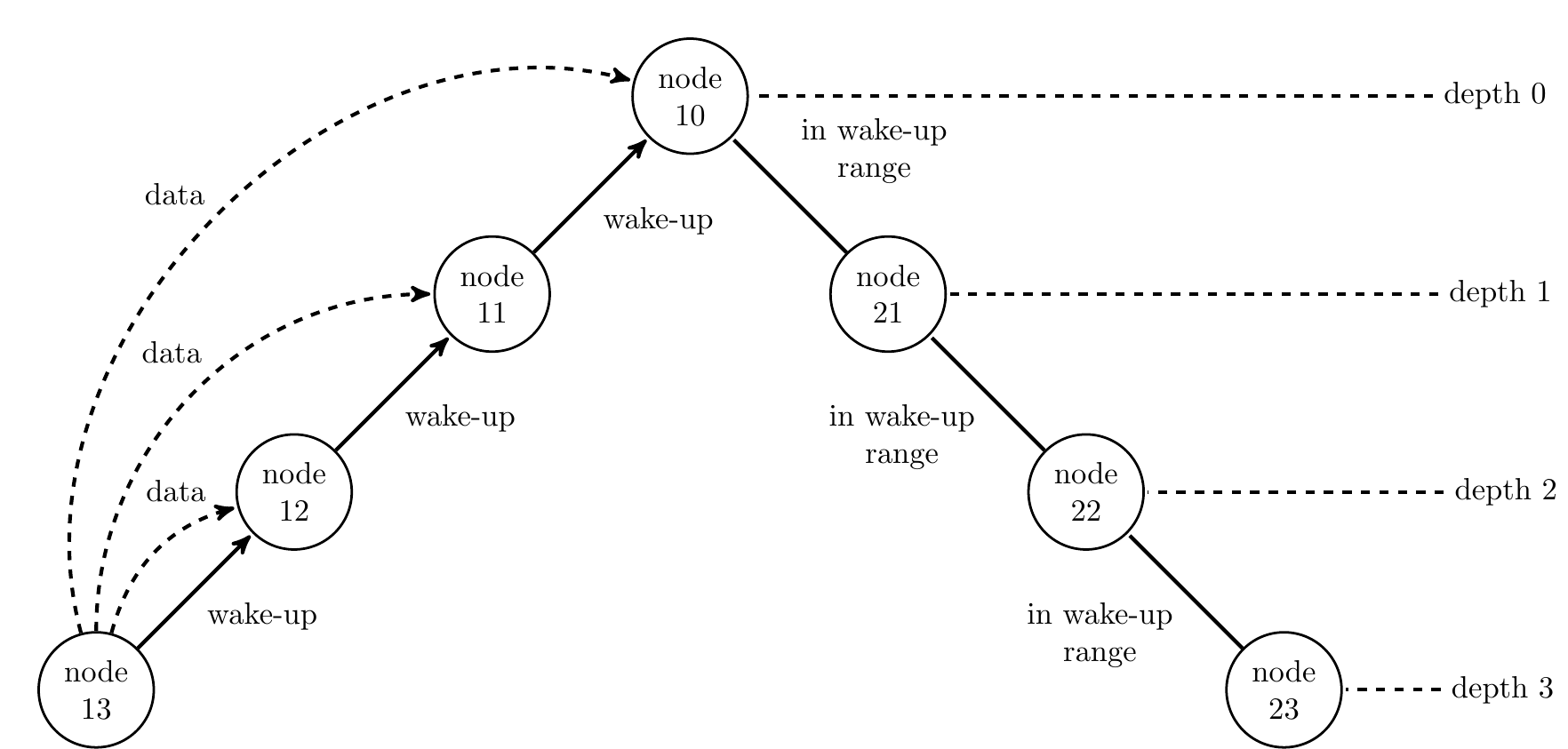}
		\caption{Schematic of the wake-up multi-hop routing protocol developed in this work.}
		\label{fig:proto}
\end{figure*}

The protocol proposed in this work is sketched in Figure \ref{fig:proto}. Sending wake-up messages is similar to the Tree Routing protocol introduced above. It is possible for nodes of depth $i$ to nodes of depth $i-1$ where they are in wake-up range. Communication data can cross several levels from depth $i$ to depth $i-n$ with $n \in \mathbb{N}$ limited by the root node and communication range. In Figure \ref{fig:proto} \textit{node 13} sends for example a wake-up signal to \textit{node 12} which forwards the wake-up to \textit{node 11} and so on until a defined maximum number $n$ of forwards or the destination is reached. Afterwards, the data can be sent directly from $node 13$ to one of the woken \textit{nodes 10}, \textit{11} or \textit{12}.

\subsection{Wake-up Layer}

\begin{figure}[hbt!]
\centering
		\includegraphics[scale=0.5]{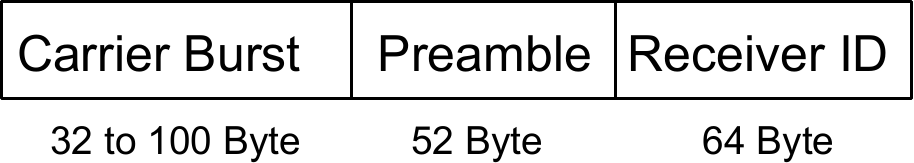}
		\caption{125 kHz wake-up call packet (WUC) including 32 to 100 byte carrier burst, 52 byte preamble and 64 byte receiver ID sent at 8192 byte per second.}
		\label{fig:wakeup_packet}
\end{figure}

\begin{figure}[hbt!]
\centering
		\includegraphics[scale=0.5]{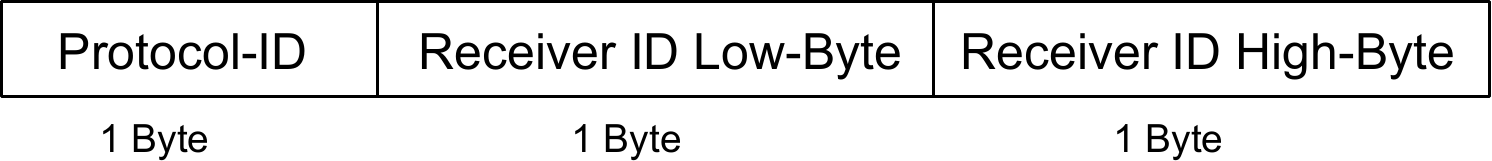}
		\caption{Wake-up acknowledge (WUC ACK) packet consisting of 3 byte (Protocol ID, Receiver ID low byte and Receiver ID high byte).}
		\label{fig:wakeup_ack}
\end{figure}

\begin{figure}[ht!]
\centering
		\includegraphics[scale=0.5]{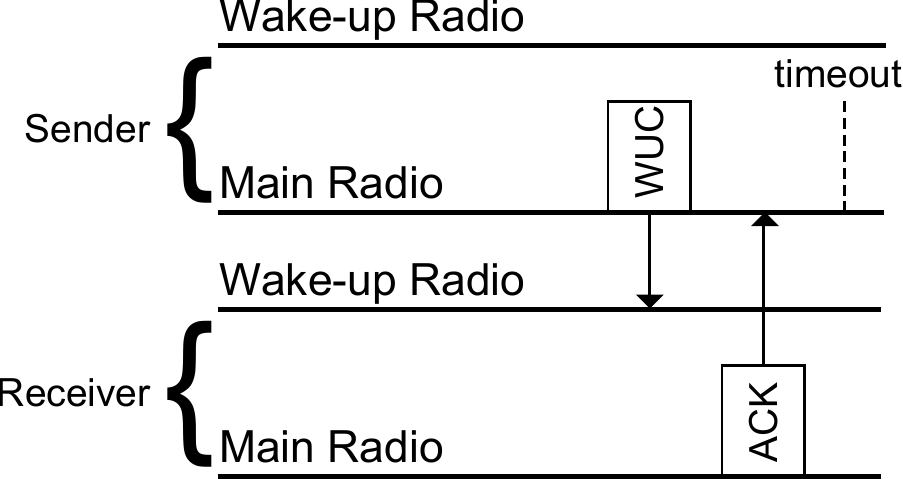}
		\caption{Packet flow of wake-up and main radios in Wake-up Layer.}
		\label{fig:wakeup_flow}
\end{figure}

The Wake-up layer is responsible for waking up of neighboring nodes. Each wake-up packet consists of carrier burst, preamble and receiver ID as depicted in Figure \ref{fig:wakeup_packet}. The carrier burst tunes the detector to the incoming frequency, the preamble is used by the detector to estimate bit length and possible offset. The receiver ID is an up to 16 bit long address to identify the receiver. When sent at a data rate of 8192 bps the wake-up message can be between 148 and 216 bytes long depending on the length of carrier burst and preamble. In a noisy environment, it is recommended to use longer carrier burst and preamble. Before an attempt is started to wake up a neighboring node each node probes the wireless channel (LBT). If a communication is currently going on, the nodes back off and restart the attempt later. After the wireless channel is found to be free each communication is initiated by sending a wake-up message. The receiver acknowledges this wake-up packet (WUC) with an acknowledge message (WUC ACK) that includes the address information of receiver and a protocol ID as can be seen in Figure \ref{fig:wakeup_ack}. If the address does not match the receiver ID or if the acknowledge message was not received before a certain timeout is reached, waking up is assumed to be unsuccessful and has to be restarted. The packet flow is schematically sketched in Figure \ref{fig:wakeup_flow}. To reduce collisions, the wake-up layer protocol realizes an RTS/CTS mechanism as depicted in Figure \ref{fig:stack}. The protocol ID is transmitted at an early stage to be able to include newer protocol versions that could react differently upon reception of certain communication packets.

\subsection{Communication MAC Layer}

\begin{figure}[ht!]
\centering
		\includegraphics[scale=0.5]{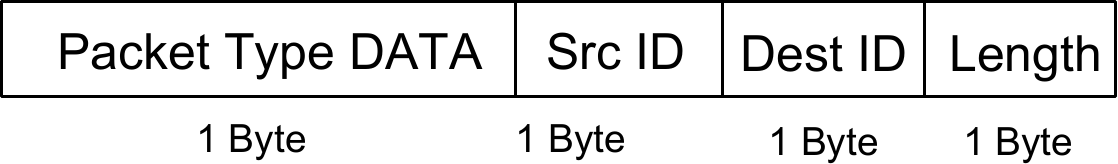}
		\caption{Data packet consisting of 4 byte (Packet Type DATA, Source ID, Destination ID and payload length).}
		\label{fig:data}
\end{figure}

\begin{figure}[ht!]
\centering
		\includegraphics[scale=0.5]{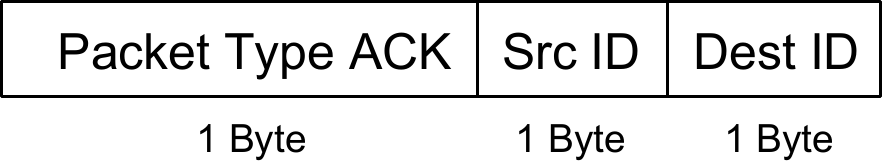}
		\caption{Acknowledge packet consisting of 3 byte (Packet Type ACK, Source ID and Destination ID).}
		\label{fig:ack}
\end{figure}

The communication MAC layer consists of two types of packets, a data packet, and an acknowledge packet. Each data packet is answered by an acknowledge packet. If the acknowledge packet is not received during a certain time frame, it is assumed that sending of data has failed. Failed data packets are reinserted into the send queue to be resent later. Figures \ref{fig:data} and \ref{fig:ack} show the data and the acknowledge packet. Packet type is used to separate the packets. IDs of the source (Src ID) and destination (Dest ID) are used to verify sender and receiver. The length byte is required internally for packet handling. 


\subsection{Routing Layer}

\begin{figure}[ht!]
\centering
		\includegraphics[scale=0.5]{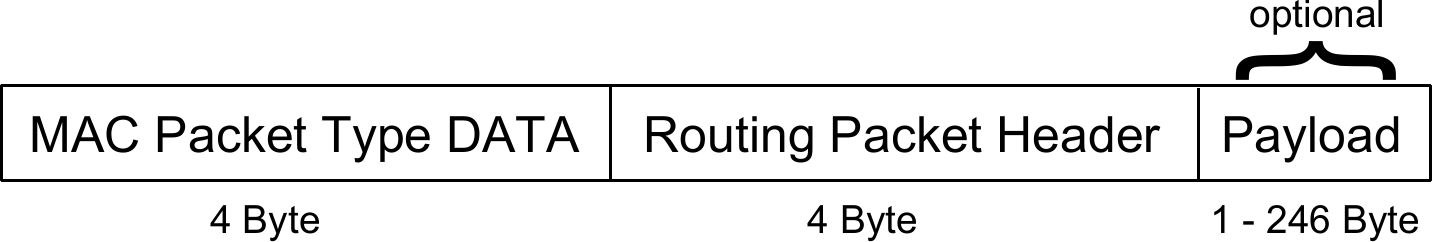}
		\caption{Routing packet embedded into MAC packet.}
		\label{fig:mac_route}
\end{figure}

While the MAC layer is responsible for the communication between neighboring nodes, the routing layer handles communications between nodes that are possibly further apart than only one hop. Routing packets are embedded into MAC layer data packets as depicted in Figure \ref{fig:mac_route}. 

\begin{figure}[ht!]
\centering
		\includegraphics[scale=0.5]{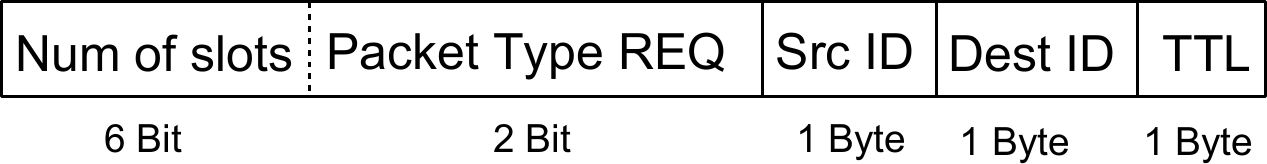}
		\caption{R\_REQ (Routing Request) packet consisting of 4 byte (Number of slots to send (6 bit) Packet Type REQ (2 bit), Source ID, Destination ID and time to live (TTL)).}
		\label{fig:routing_req}
\end{figure}

\begin{figure}[ht!]
\centering
		\includegraphics[scale=0.5]{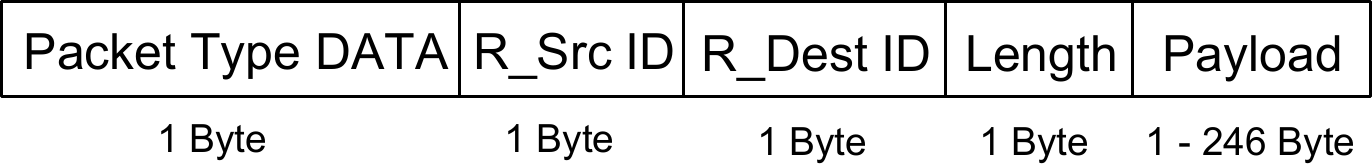}
		\caption{DATA (Routing Data) packet consisting of 4 byte (Packet Type DATA, Routing Source ID, Routing Destination ID and payload length).}
		\label{fig:routing_data}
\end{figure}

\begin{figure}[ht!]
\centering
		\includegraphics[scale=0.5]{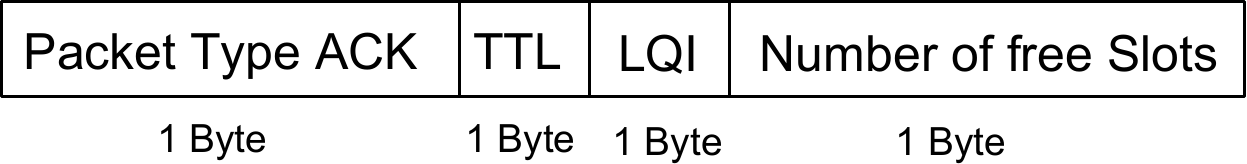}
		\caption{R\_REQ\_ACK (Routing Acknowledge) packet consisting of 4 byte (Packet Type ACK, current time to live (TTL), Link Quality Identifier (LQI) and number of available memory slots).}
		\label{fig:routing_ack}
\end{figure}

The routing layer takes care of sending, receiving and forwarding packets from source to destination. Figures \ref{fig:routing_req}, \ref{fig:routing_data} and \ref{fig:routing_ack} show the three available routing packets, namely routing request (R\_REQ), data (DATA) and acknowledge (R\_REQ\_ACK). Each packet consists of four bytes. All data to be sent is managed in data slots that form the message queue. The first six bits of a request type packet are reserved for the number of slots to be sent in the currently ongoing communication. R\_Src Id and R\_Dest ID are the routing source and destination IDs of the communicating nodes which could be equal to the MAC IDs but can also be different. TTL (time to live) indicates how many hops a request can be forwarded. Upon reception of a routing request, the receiving node decreases TTL by one, before forwarding the request to the next node. In case TTL is zero the request will not be further forwarded. Forwarding of routing requests is realized with route request packet type packets keeping source and destination ID untouched.


\begin{figure}[ht!]
\centering
		\includegraphics[scale=0.15]{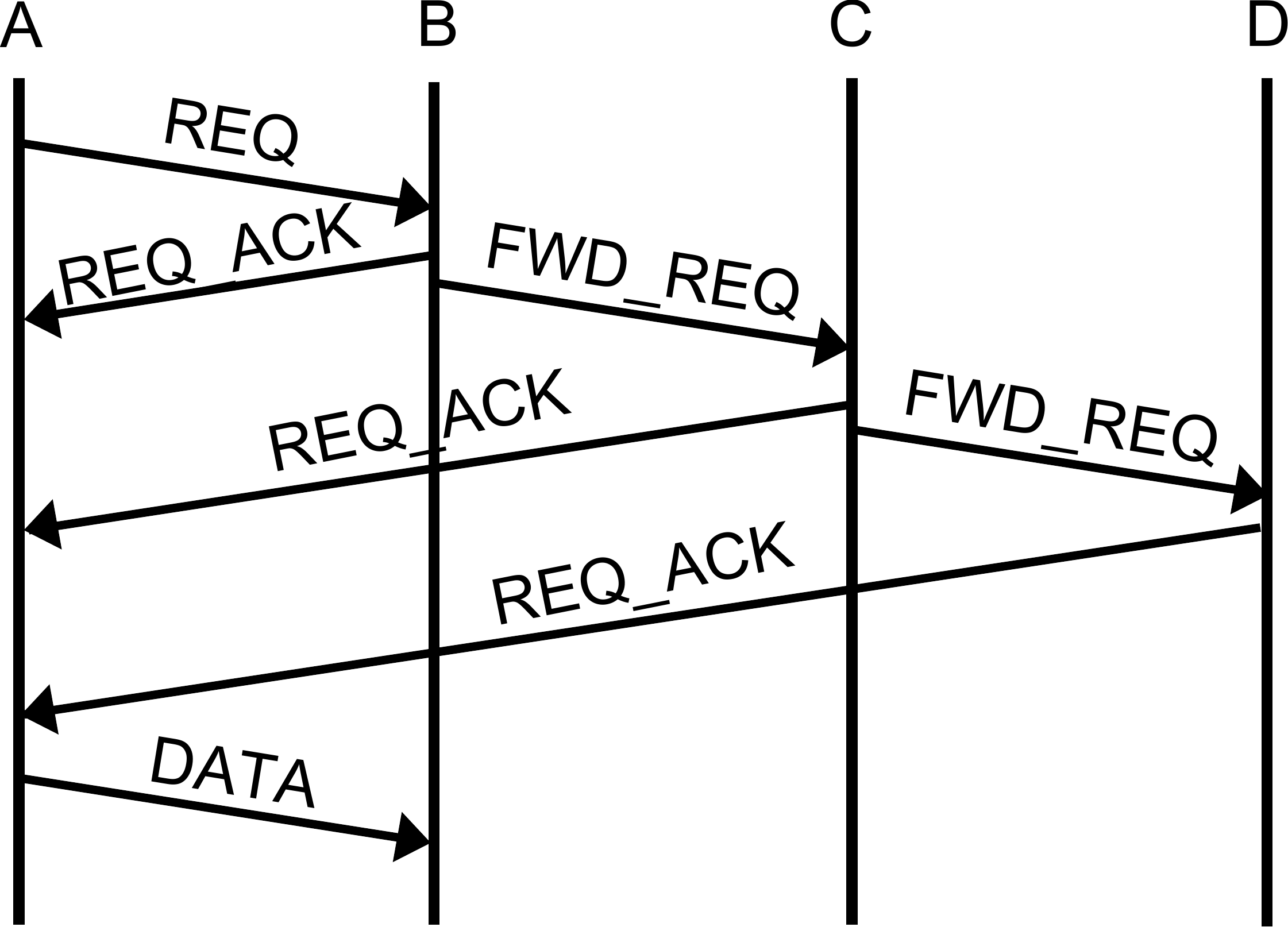}
		\caption{Sequence diagram of the routing protocol in case the data is sent to the next neighbor. Decision to where the data are sent is done at \textit{node A} based on information included in the request acknowledge data.}
		\label{fig:caseA}
\end{figure}

\begin{figure}[ht!]
\centering
		\includegraphics[scale=0.15]{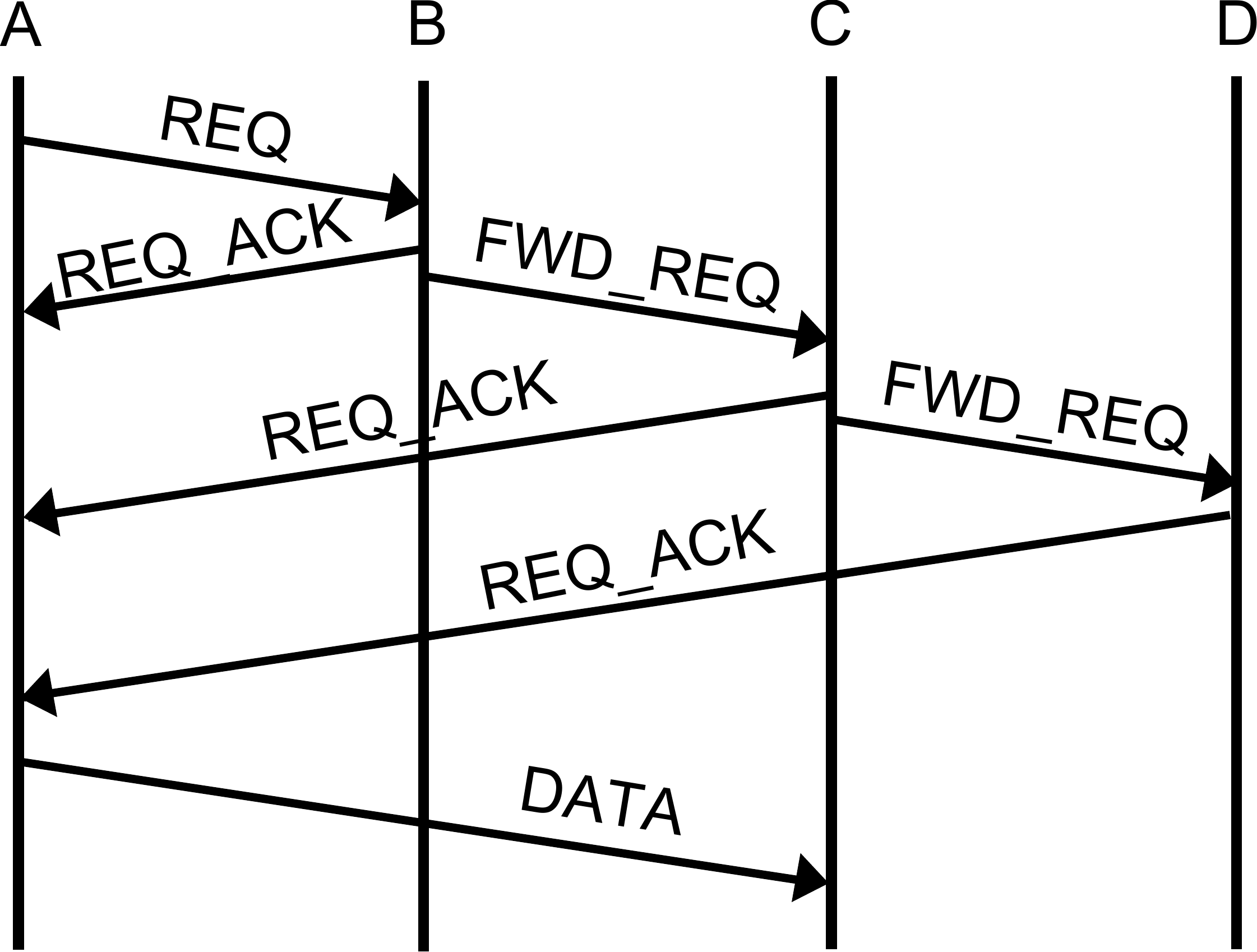}
		\caption{Sequence diagram of the routing protocol for communication to the two-hop distant neighbor. Decision to where the data are sent is done at \textit{node A} based on information included in the request acknowledge data.}
		\label{fig:caseB}
\end{figure}

\begin{figure}[ht!]
\centering
		\includegraphics[scale=0.15]{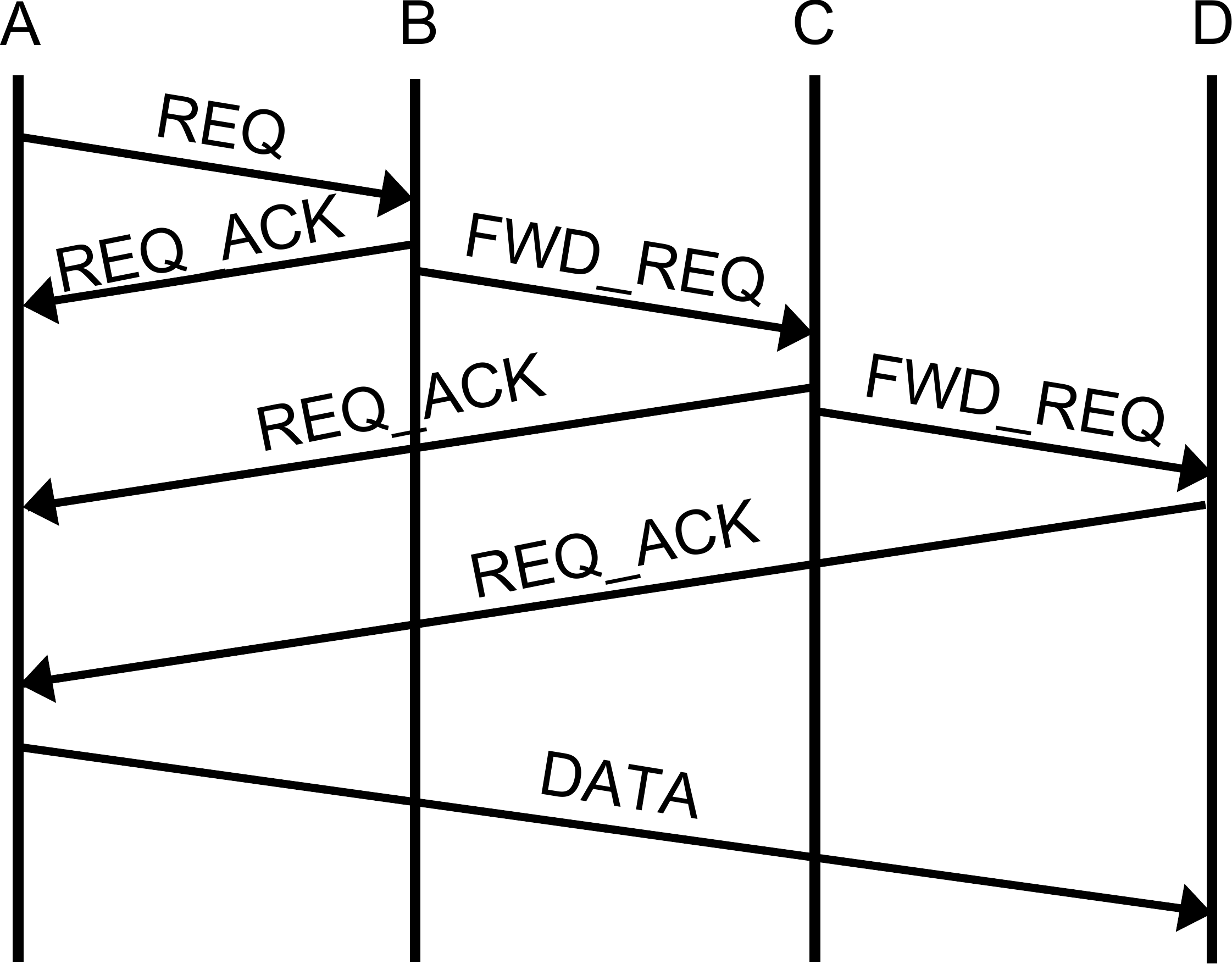}
		\caption{Sequence diagram of the routing protocol for communication to a three-hop distant neighbor. Decision to where the data are sent is done at \textit{node A} based on information included in the request acknowledge data.}
		\label{fig:caseC}
\end{figure}

Figures \ref{fig:caseA}, \ref{fig:caseB} and \ref{fig:caseC} show the sequence diagrams of the routing protocol in case of four participating nodes. \textit{Node A} is the source node, \textit{nodes B} and \textit{C} are possible relay nodes and \textit{node D} is the sink. \textit{Node A} starts by sending a routing request (R\_REQ) to \textit{node B}. \textit{Node B} forwards the request (FWD\_REQ) to its next neighbor \textit{node C} who will again forward the request to \textit{node D}. Each node (B, C, and D) answers the request by sending of a request acknowledge (R\_REQ\_ACK) to \textit{node A}. \textit{Node A} collects all request acknowledgments and decides based on the information included in the acknowledges to which node the data will be sent. Currently implemented parameters that support the decision, to which node data is sent to, are: available data slots at the receiving node and hop distance from starting node. Further parameters like the available energy at receiver node or various status data like link quality or number of successful wake-ups can be easily used to increase the network stability.


Once a communication link to a node is established, up to 64 data packets consisting of up to 246 bytes each can be transmitted in a row. After transmission, the link gets closed and the participating nodes fall back to sleep, again. The same routing scheme is repeated until all data has reached their destination.

\subsection{State Machine}

\begin{figure}[ht!]
\centering
		\includegraphics[scale = 0.5]{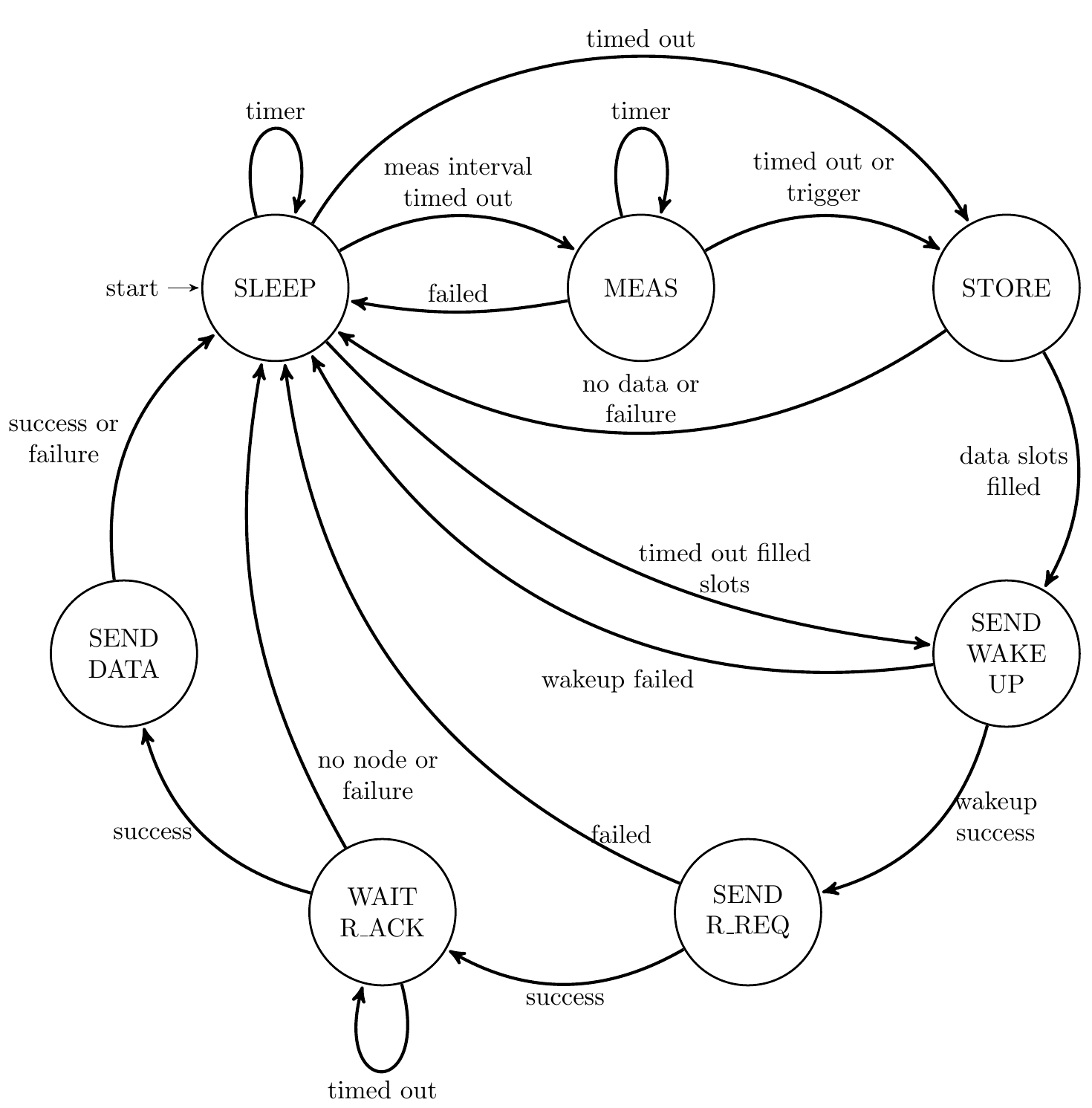}
		\caption{State machine of a sensor node for data transmission.}
		\label{fig:sm_sender}
\end{figure}

As introduced in \cite{kumbergPLSE}, the embedded software is implemented as a state machine as depicted in Figure \ref{fig:sm_sender}. At the beginning, a sensor node is in SLEEP state in which it consumes only minimal energy. A low-energy timer transfers the node from sleep either to start a sensor measurement (state MEAS) or to check if there is data available in the memory that is not yet sent (state STORE).

In case there are already prepared data slots available, for example from a previously aborted sending, the sleep state will be left and data transfer is initiated by sending a wake-up signal (state SEND WAKE-UP). After a measurement, sensor data is stored in a ringbuffer on the microSD card and data packets are prepared and moved into one of up to 64 available data slots. If there are no free slots available the data is kept in memory to be processed later.

After successful filling the message queue, sending of data is initiated with a wake-up signal (state SEND WAKE-UP). Successfully waking of the neighbor node, is indicated by a wake-up acknowledge and a routing request is sent (state SEND R\_REQ) containing destination ID, number of data packets and max number of wake-up hops. Then, the node listens for route request acknowledgments sent by the woken nodes (state WAIT R\_ACK). If at least one node that answers has a free slot available, the node starts to send all possible data packets (state SEND DATA). After successful sending, or if any error occurs, the node exits its current state and goes back to sleep. The state machine of the receiver is similar to that of the transmitter.

Looking at the state machine, it becomes clear that in case two sensor nodes try to send data at the same time, the data packets would collide and packet transmission would fail. Additionally, T-ROME can encounter self-interference due to the forwarding mechanism of packets that are sent at the same time. To avoid collisions, each source node (but not the relay nodes, as the channel is assumed to be busy during the complete period of data transmission) probes the wireless channel before transmission and if it finds the channel busy it backs off for a certain time period before testing the channel again. To calculate the back off period a simple algorithm is used that calculates the back-off time based on the unique node ids. This means that nodes further away from the sink node have longer back off periods than nodes nearer to the sink. This avoids self-interference and reduces congestions near the sink during periods of high data traffic.

\subsection{Wake-up Message}
\label{sec:wakeup_message}

The wake-up signal was received at a data rate of 8192 kbit per second (bit length: 122 \textmu s), which means a 125 kHz period requires 4 byte ones and 4 byte zeros sent in a row at 250 kbit per second, resulting in a bit length of 128 \textmu s. 
From sender (receiver) side, the wake-up message consisted of 42 byte (10 bit) Carrier Burst which is required at the receiver to detect the presence of a signal and to fine-tune its internal frequency to the incoming signal frequency. The preamble consisted of 48 byte (12 bit). Its purpose is to adjust the receiver offset to be approximately at the level of the averaged input signal and to verify the bit length. The pattern depicts the 16 bit address of the wake-up receiver.  It requires sending of 64 byte (16 bit). In the case of Manchester coding, this results in an 8 bit address that can be used to address up to 256 independent devices. For example, the node ID sent in Figure \ref{fig:wakeup_signal} is decimal 85.

\begin{figure}[ht!]
\centering
		\includegraphics[scale=0.5]{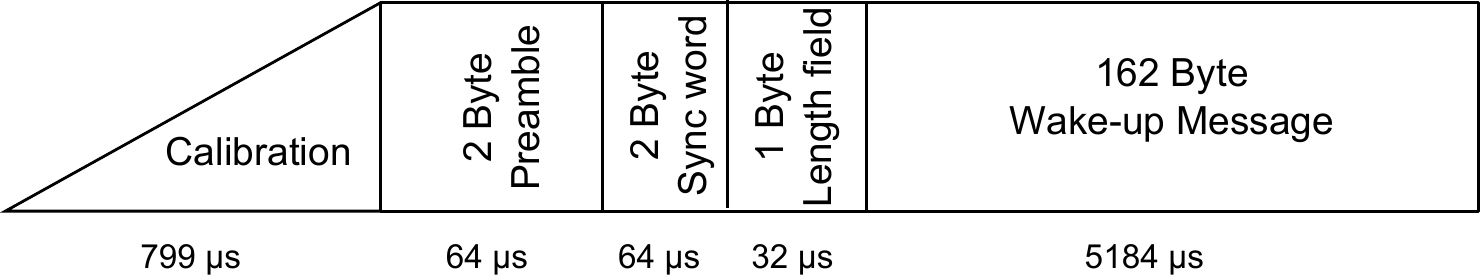}
		\caption{Complete wake-up packet including calibration and mandatory radio bytes.}
		\label{fig:radio_wakeup_packet}
\end{figure}

Figure \ref{fig:radio_wakeup_packet} shows the message schematically. Before sending, the radio requires a calibration cycle. Preamble, a sync word, and length field are mandatory bytes which make a wake-up message 6143 \textmu s long. Out of that, the radio is for 5344 \textmu s in sending state and 799 \textmu s in calibration state.

\subsection{Communication Packets}
\label{sec:com_pack}
\begin{figure}[ht!]
\centering
		\includegraphics[scale=0.42]{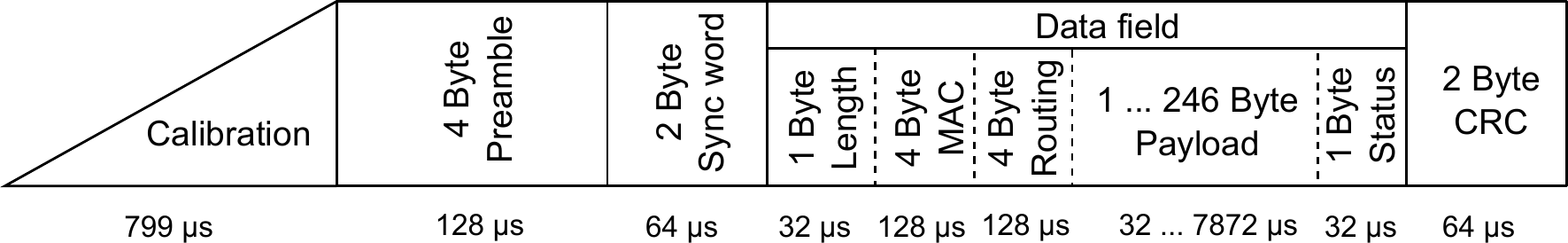}
		\caption{Radio packet including calibration.}
		\label{fig:radio_packet}
\end{figure}

Figure \ref{fig:radio_packet} shows schematically the buildup of a complete radio packet including calibration of the radio, sending of the preamble, sync word, length, MAC, status and CRC bytes. Sending of payload and routing bytes is optional. All times (including calibration) in Figure \ref{fig:radio_packet} are calculated for a baud rate of 250 kbit per second and GFSK (gaussian frequency shift keying) modulation. Generally, sending of data is separated into hardware specific and protocol layer specific parts. In sum, each packet requires the hardware specific calibration, preamble, sync word and CRC which add up to around 991 \textmu s. The rest of the time is required to send protocol messages, either wake-up, MAC or routing. A MAC packet requires 1247 \textmu s and a routing packet without payload requires 1375 \textmu s. The payload is sent in additionally 32 - 7872 \textmu s, depending on payload size. According to the datasheet, the radio draws around 8.4 mA during calibration and when sending at 868 MHz, 0 dBm gain around 16.4 mA. In receive state, the radio requires around 16.9 mA and for sending a wake-up call at +12 dBm gain the CC1101 draws 34.2 mA.

\section{Experimental Analysis}
\label{sec:exp_ana}
In order to verify the assumptions on current consumption and timing intervals (as discussed in Section \ref{sec:network_protocol}), we analyzed a wake-up message, communication messages and the protocol on the whole, experimentally. The results are presented in following Sections \ref{sec:ana_wakeup} and \ref{sec:ana_comm}. Furthermore, we experimentally investigated the occurrence of false positive and false negative wake-ups as introduced in Section \ref{sec:false_wakeup}.

\subsection{Wake-up Message}
\label{sec:ana_wakeup}
\begin{figure}[ht!]
\centering
		\includegraphics[scale=0.18]{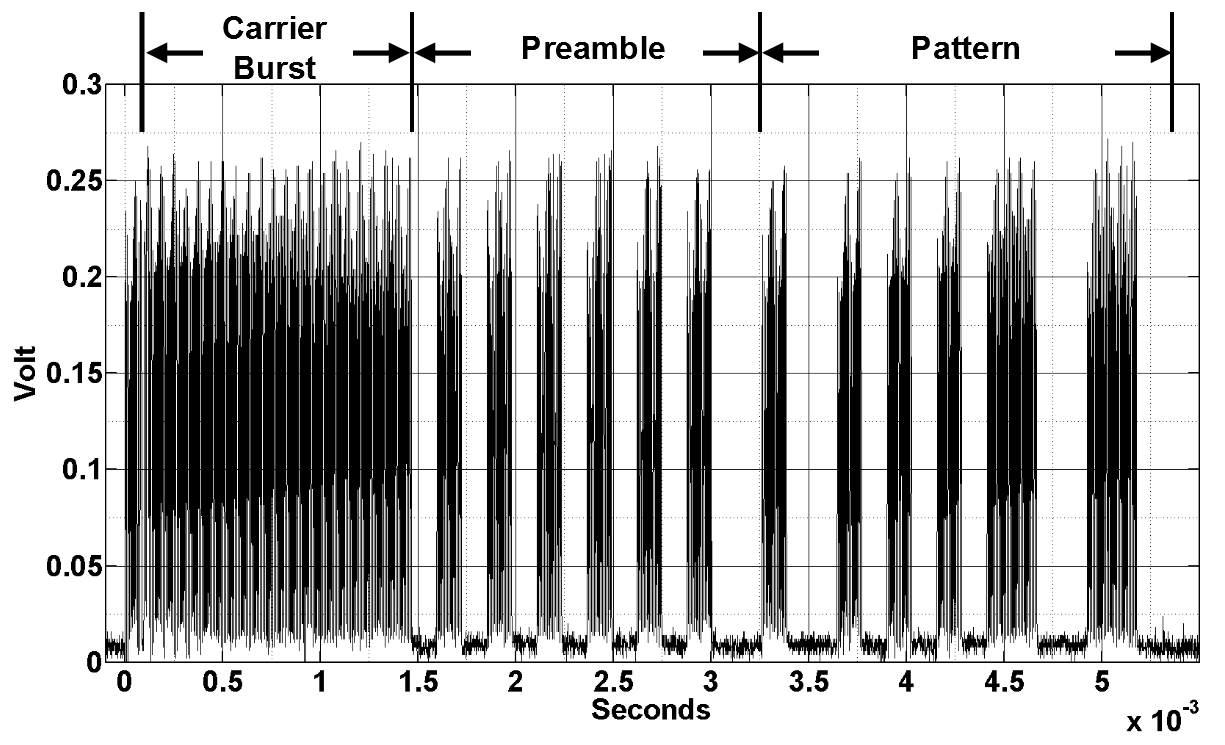}
		\caption{Manchester coded wake-up signal consisting of Carrier Burst, Preamble and address Pattern.}
		\label{fig:wakeup_signal}
\end{figure}

Figure \ref{fig:wakeup_signal} shows the wake-up message used in this work captured at the output of the envelope detector. The wake-up message was Manchester encoded to improve stability and to reduce the false wake-up rate as introduced in Section \ref{sec:false_wakeup}. Manchester encoding, in this case, means that a binary one results from a transition from high to low, and a binary zero results from the transition from low to high. So one bit Manchester encoded requires two bit sent. 

Figure \ref{fig:wakeup_signal} clearly shows that the length of the real wake-up message corresponds very well to the theoretical length of the wake-up message calculated by using the numbers provided by the datasheets as given in Section \ref{sec:wakeup_message}. 

\subsection{Communication Messages}
\label{sec:ana_comm}
\begin{figure}[ht!]
\centering
		\includegraphics[scale=0.13]{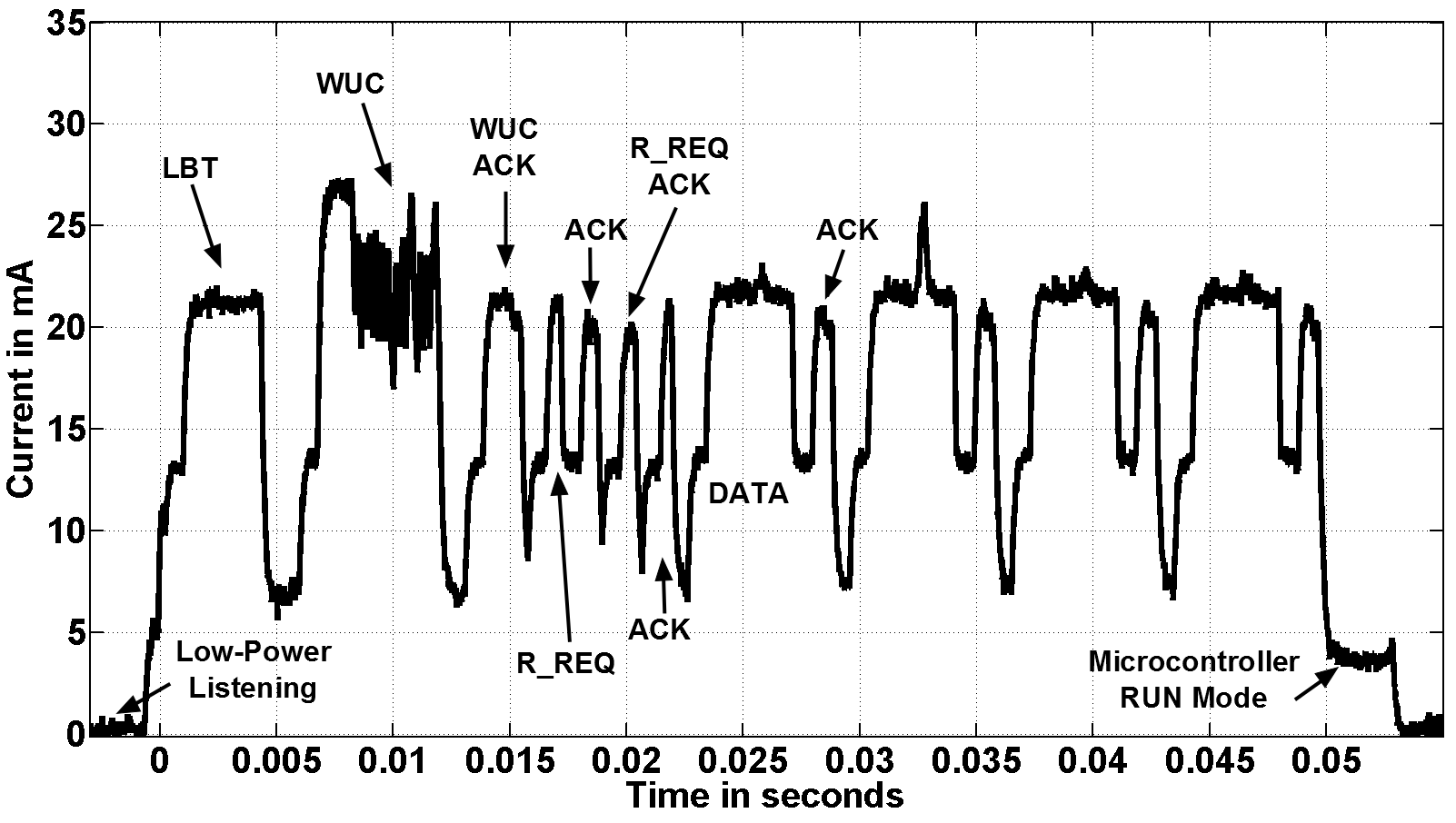}
		\caption{Current drawn by the sensor node in the different states of the protocol.}
		\label{fig:current_consumption}
\end{figure}

Figure \ref{fig:current_consumption} shows exemplary the current consumption of a sensor node in the different states of the proposed protocol measured via a shunt resistor in the power line. In this example, the node sent 4 data packets consisting of 100 byte each to the next neighbor node. It can be seen that the currents provided in Sections \ref{sec:sensor_node} and \ref{sec:com_pack} for microcontroller and radio fit very well to the measurement results for radio calibration, sending and receiving of communication packets, low-power listening, and microcontroller run mode current. It can be further seen, that sending of wake-up packets require less current than expected from the datasheet numbers, only. This is due to the fact that the Manchester encoded wake-up packets consist of an equal amount of zeros and ones and the radio power is reduced during sending of zeros. Furthermore, we can see that the timing fits very well to the suggested timing calculated in Sections \ref{sec:wakeup_message} and \ref{sec:com_pack}.

We used a logic analyzer to visualize all sending and receiving states. As laboratory test setup we used the same configuration as introduced in Figure \ref{fig:proto} with four participating nodes: \textit{node 13} as source, \textit{node 12} and \textit{node 11} as relay nodes and \textit{node 10} as sink. Node 13 sent 5 data packets of 100 bytes payload each. According to the protocol \textit{nodes 12} and \textit{11} forwarded the request to \textit{node 10} that finally received all data packets after around 90 ms.

\begin{figure*}[ht!]
\centering
		\includegraphics[scale=0.5]{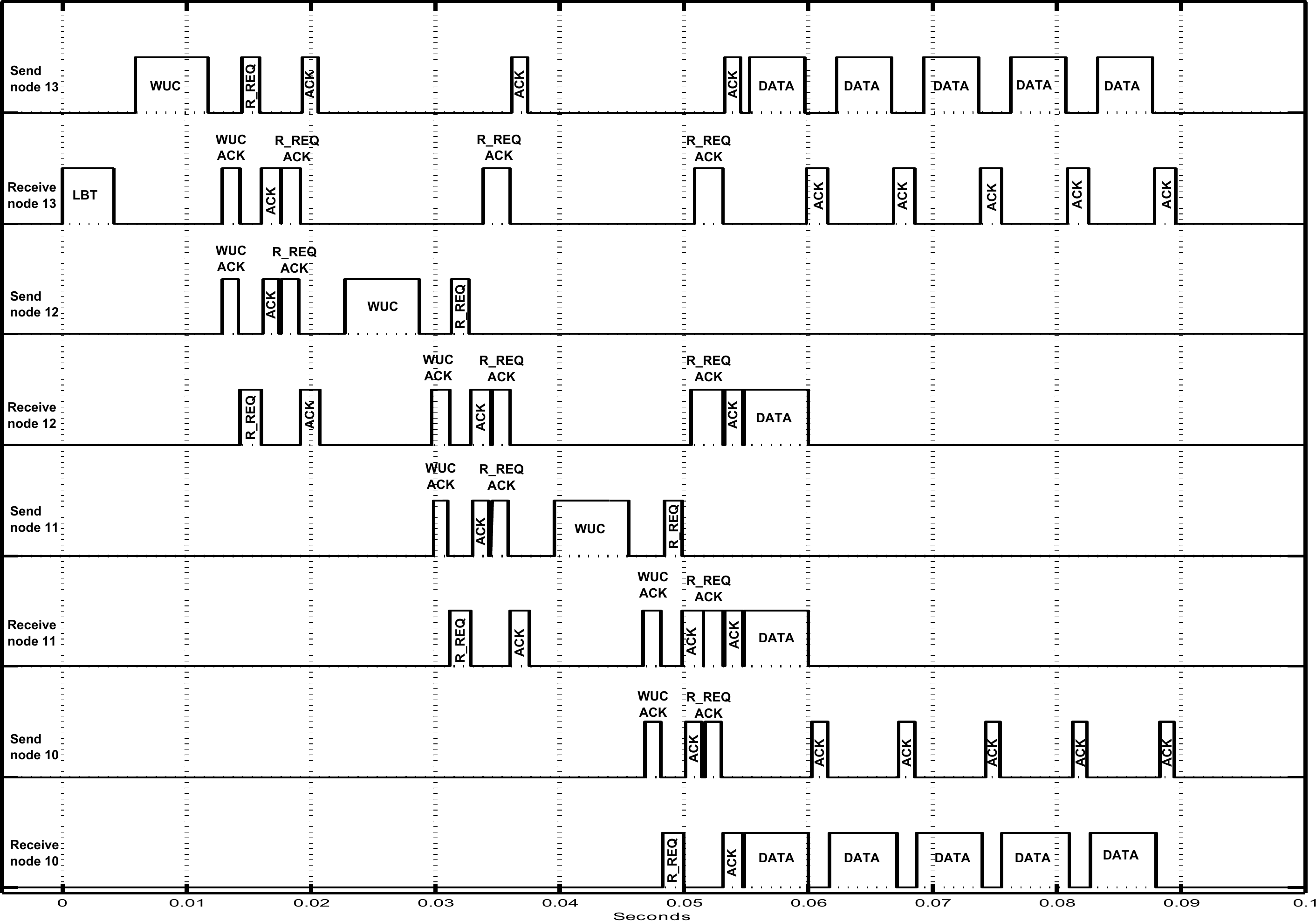}
		\caption{Flow chart of the protocol for sending and receiving of 5 data packets in case of 4 participating nodes. Each node (node 13, node 12, node 11 and node 10) has a sending (upper line) and receiving (lower line) state. Node 13 is source, node 10 is sink. Nodes 12 and 11 forward the wake-up calls.}
		\label{fig:routing_states}
\end{figure*}

These times intervals can be seen in Figure \ref{fig:routing_states} which shows the sending of 500 bytes payload in 5 packets of 100 bytes each over a row of four nodes as sketched in Section \ref{sec:network_protocol} Figure \ref{fig:proto}.

\subsection{False Positive and False Negative Wake-ups}

Other factors to be taken into account are false positive and false negative wake-ups, as shown in Section \ref{sec:false_wakeup}. To evaluate the occurrence of false negative wake-ups we conducted two laboratory experiments consisting of a sender and a receiver, first connected by cables and second by antennas. The first setup was chosen to easily place an active attenuator in line to be able to reduce the incoming signal from 0 dBm to -60 dBm. The second test was chosen to verify that external interferences have no influences on the wake-up rate. 

\begin{figure}[ht!]
\centering
		\includegraphics[scale = 0.16]{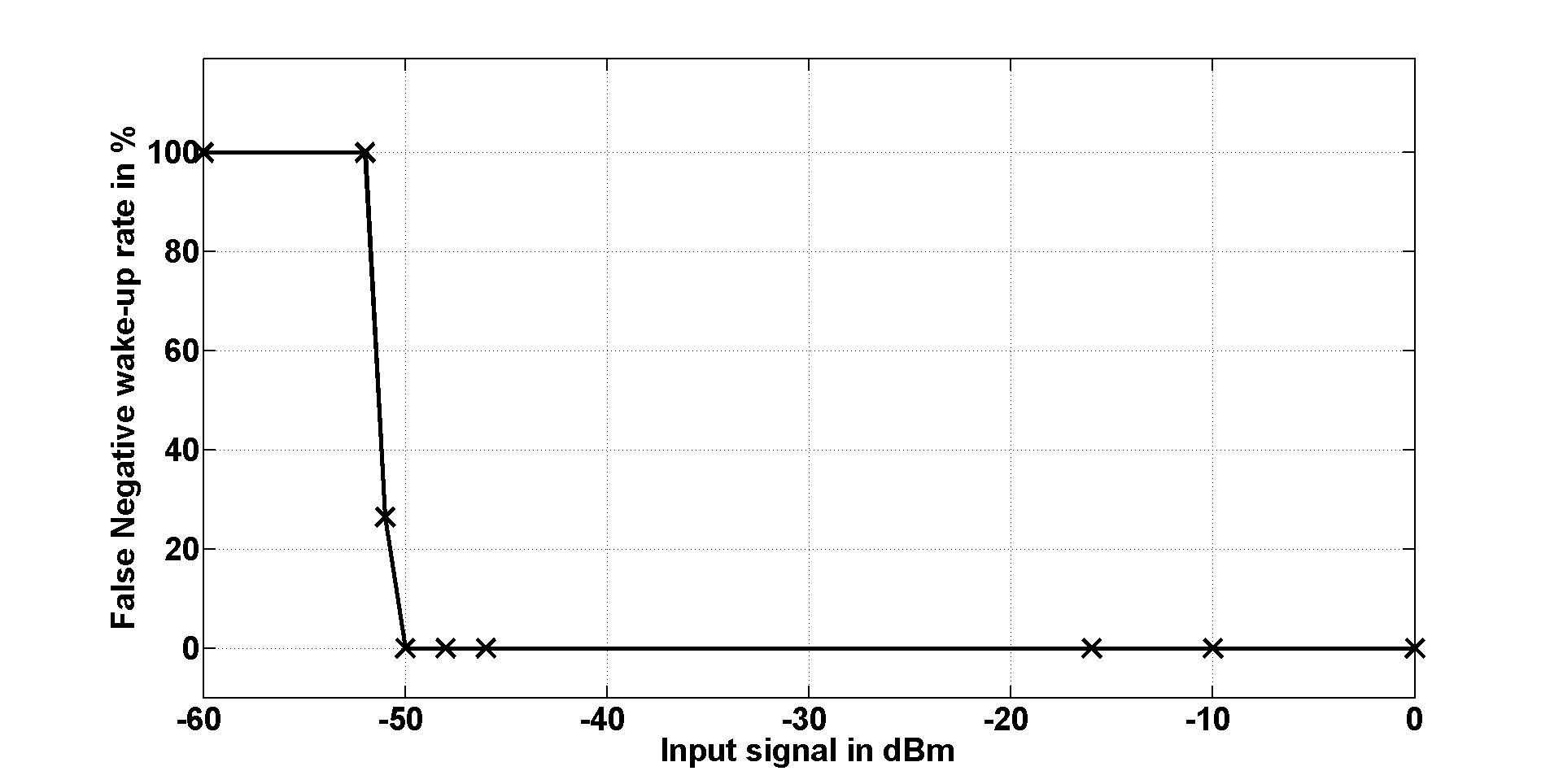}
		\caption{Experimentally measured false negative wake-up rate over input signal strength in dBm, nodes connected by cables.}
		\label{fig:false_wakeups}
\end{figure}

During both tests, the sender sent each possible address (0x00 to 0xFF) 100 times. After sending 100 addresses the sender signaled the receiver via a separate connection. After receiving this signal the receiver incremented its address stepwise from 0x00 to 0xFF. Figure \ref{fig:false_wakeups} shows the averaged false wake-up rate over input signal strength for the nodes connected by cables. The experiment shows that the wake-up receiver has no false negative wake-ups until the signal strength reaches its sensitivity limit at around -50 dBm. Then, the false negative wake-up rate increases quickly to 100 \% for signals sent below -52 dBm. 

\begin{figure}[ht!]
\centering
		\includegraphics[scale = 0.15]{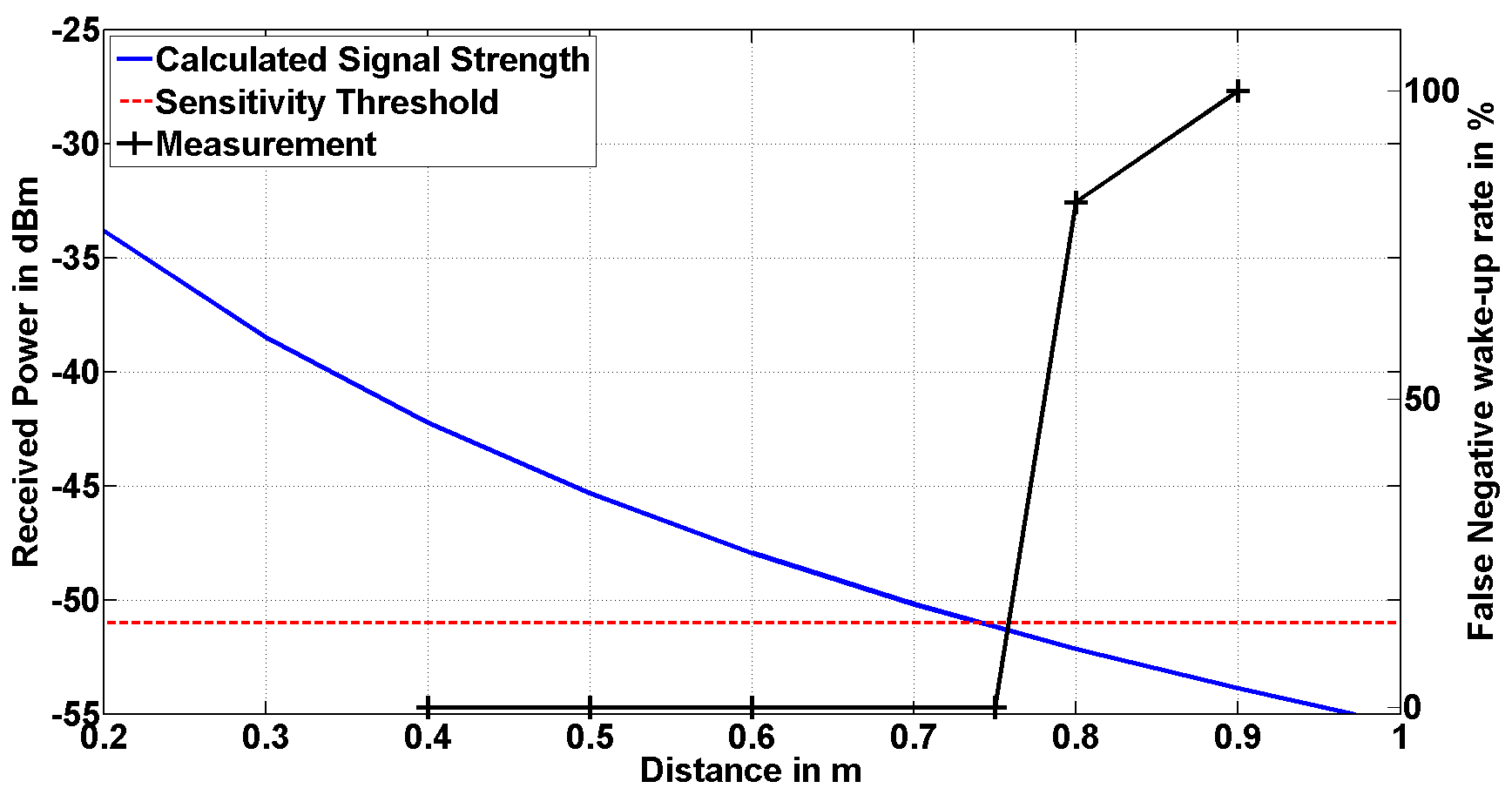}
		\caption{Experimentally measured false negative wake-up rate over distance for the nodes connected by antennas on the right axis and calculated signal strength over distance on the left axis. The dashed line shows the sensitivity level of the receiver at around -51 dBm.}
		\label{fig:false_wakeups_exp_2}
\end{figure}

Figure \ref{fig:false_wakeups_exp_2} shows the averaged false negative wake-up rate over distance for the nodes connected by antennas on the right axis and calculated signal strength over distance on the left axis. The signal strength was calculated using the equations given in \cite{kvaksrud2008range}. The dashed line shows the sensitivity level of the receiver at around -51 dBm. This experiment was conducted indoors and sender power was set to -20 dBm to ensure short wake-up ranges and to be able to use the signaling connection between sender and receiver. Both experiments show similar results and the receiver has no false negative wake-ups until the signal strength reaches its sensitivity limit at around -50 dBm to -51 dBm and then increases quickly to 100 \% for signals sent below -52 dBm. 

To analyze the occurrence of false positive wake-ups we conducted an experiment similar to the one described above. We used the same test setup and  configured the sender to send every address from 0x00 to 0xFF 10 times, while the receiver kept its address. Only each time the sender sent 0xFF the receiver incremented its address by one. Since the receiver's initial address was 0x00, each address could be cross-checked with all other possible addresses during this test. The receiver just woke up 10 times, exactly what would be expected if no false positive wakeups occur. Throughout the test, the received signal strength was set to -25 dBm.
  
\section{Numerical Analysis}
\label{sec:analysis}
In order to analyze the performance of our proposed protocol, and to compare it to other protocols, we introduce a Markov chain based model and meta-models of the routing algorithms T-ROME, CTP-WUR and of an algorithm that we here call the naive algorithm. We decided to compare our protocol especially to these two protocols, as most implemented networks use some derivate of the naive algorithm or a relaying mechanism similar to CTP-WUR. 

States of the meta-models can either be $w$ or $T$. States $w$ are states where a node attempts to wake-up another node and states $T$ depict states in which data should be transferred from one node to another node. The models consist of a row of $m$ nodes, subscribed with $j$ and $i$, where $i<j$.

To model the algorithms, we further assume a message existent at node $i$ at time $t_0$. Then, a $w$ state could for example be, $w_{i,i,i+1}$, which means that the message is at node $i$ (first subscript) and node $i$ (second subscript) attempts to wake up the next node $i+1$ (third subscript). In the case of $T$ states the subscripts have the following meaning: $T_{i,j}$ means a data transmission from node $i$ to node $j$. Transitions between states are possible along the arrows which are connected to a certain cost that can be probability or time in a more general way. We use subscript $q$ to describe the probability of a successful wake-up, and $p$ to describe the probability of a successful data communication. To simplify the models we assume equal success probabilities for all nodes, that is $\forall i \in m: p_i = p, q_i = q$.

Our Markov chain based model reflects errors on the medium access level and does not describe the dynamic routing behavior originating from changes in link quality estimations or due to changes in the energy level of certain nodes that could lead to different routes. This could potentially lead to different behaviors of the routing algorithms, and the comparisons presented in Section \ref{sec:results} might be influenced by this. An extended Markov chain based model that also reflects the dynamic behavior is clearly more complex and may be part of our future research.

\subsection{T-ROME}
\begin{figure}[ht!]
\centering
		\includegraphics[scale = 0.8]{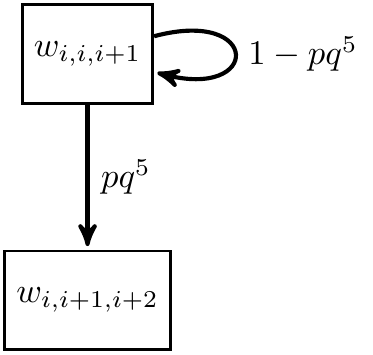}
		\caption{T-ROME meta model for node $i$ attempting to wake-up node $i+1$. The message is at node $i$.}
		\label{fig:t-rome_meta_model_1}
\end{figure}

\begin{figure}[ht!]
\centering
		\includegraphics[scale = 0.8]{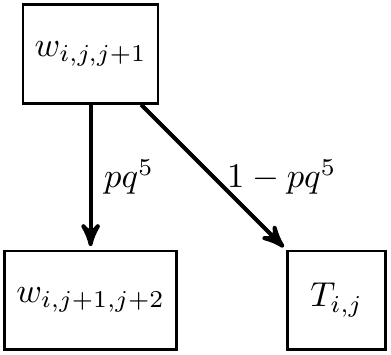}
		\caption{T-ROME meta model for node $j$ attempting to wake-up node $j+1$. The message is at node $i$.}
		\label{fig:t-rome_meta_model_2}
\end{figure}

\begin{figure}[ht!]
\centering
		\includegraphics[scale = 0.8]{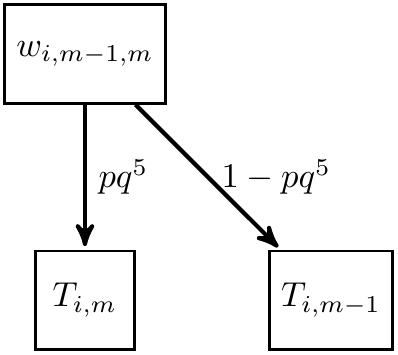}
		\caption{T-ROME meta model for node $m-1$ attempting to wake-up node $m$. The message is at node $i$.}
		\label{fig:t-rome_meta_model_3}
\end{figure}

\begin{figure}[ht!]
\centering
		\includegraphics[scale =0.8]{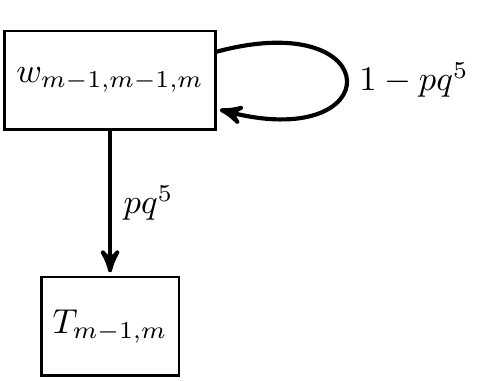}
		\caption{T-ROME meta model for node $m-1$ attempting to wake-up node $m$. The message is at node $m-1$.}
		\label{fig:t-rome_meta_model_4}
\end{figure}

\begin{figure}[ht!]
\centering
		\includegraphics[scale = 0.8]{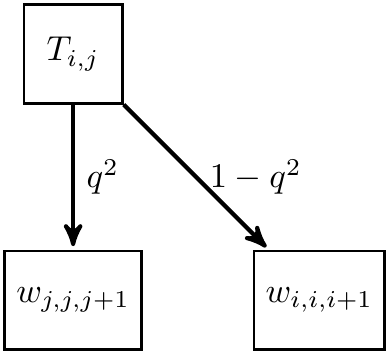}
		\caption{T-ROME meta model for node $i$ attempting to transmit data to node $j$.}
		\label{fig:t-rome_meta_model_5}
\end{figure}

\begin{figure}[ht!]
\centering
		\includegraphics[scale = 0.8]{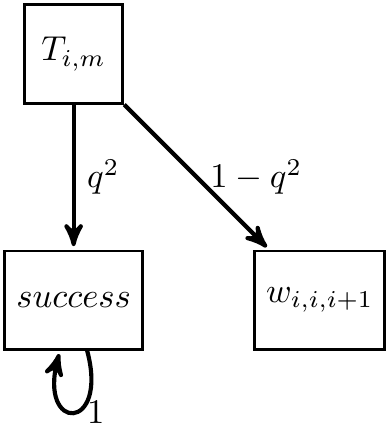}
		\caption{T-ROME meta model for node $i$ attempting to transmit data to node $m$.}
		\label{fig:t-rome_meta_model_6}
\end{figure}

T-ROME, wake-up messages can be forwarded or send directly. In summary, there exist following four possible meta states for a T-ROME branch consisting of $m$ nodes:
\begin{itemize}
\item The message is at node $i$ and node $i$ tries to wake-up node $i+1$, for $i<m-1$. When awake, node $i+1$ tries to wake-up node $i+2$. The message is still at node $i$. If wake-up of node $i+1$ fails, the message stays at node $i$ that will initiate another wake-up attempt at a later time. Figure \ref{fig:t-rome_meta_model_1} depicts this case.
\item The message is at node $i$ and node $j$ tries to wake-up node $j+1$, for $i<j$ and $j+1 < m$. When awake, node $j+1$ tries to wake-up node $j+2$. The message is still at node $i$. If wake-up of node $j+1$ fails, node $j$ is ready to receive the message. Figure \ref{fig:t-rome_meta_model_2} depicts this case.
\item The message is at node $i$ and node $m-1$ tries to wake-up node $m$, for $i<m-1$. After reception of the wake-up message, node $m$ is ready to receive the message from node $i$. If wake-up of node $m$ fails, node $m-1$ is ready to receive the message. Figure \ref{fig:t-rome_meta_model_3} depicts this case.
\item The message is at node $m-1$ and node $m-1$ tries to wake-up node $m$. After successfully waking up node $m$ it is ready to receive the message from node $m-1$. If wake-up of node $m$ fails, the message stays at node $m-1$ that will initiate another wake-up attempt at a later time. Figure \ref{fig:t-rome_meta_model_4} depicts this case.
\end{itemize}
In case of data transmission, there exist following two possibilities:
\begin{itemize}
\item Node $i$ tries to transmit the message to node $j$, for $i<j<m$. If it succeeds, node $j$ has the message and tries to wake-up node $j+1$. If it fails, the message stays at node $i$ that will initiate another wake-up attempt at a later time. Figure \ref{fig:t-rome_meta_model_5} depicts this case.
\item Node $i$ tries to transmit the message to node $m$, for $i<m$. If it succeeds, node $m$ has the message. If the transmission fails, the message stays at node $i$ that will initiate another wake-up attempt at a later time. Figure \ref{fig:t-rome_meta_model_6} depicts this case.
\end{itemize}

\begin{figure}[ht!]
\centering
		\includegraphics[scale = 0.8]{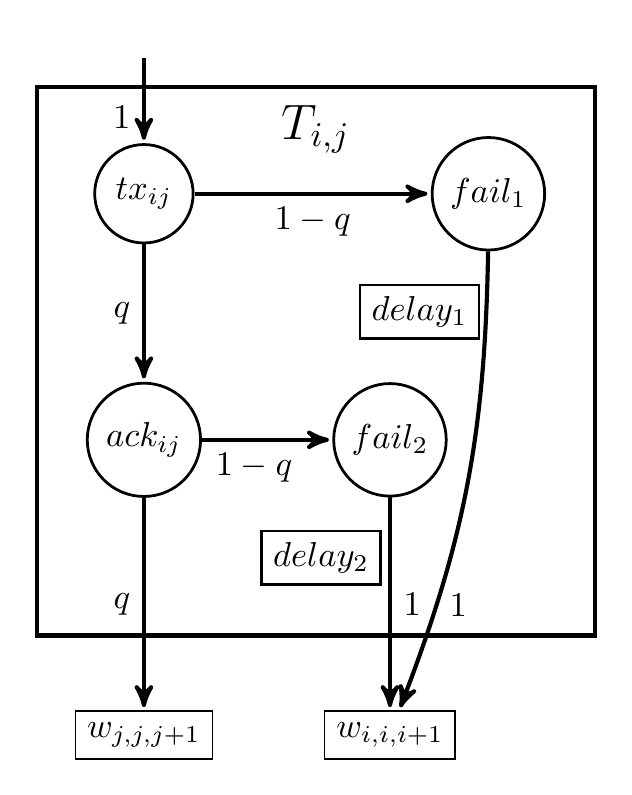}
		\caption{Markov chain for node $i$ attempting to send data to node $j$}
		\label{fig:markov_tij}
\end{figure}

\begin{figure}[ht!]
\centering
		\includegraphics[scale = 0.8]{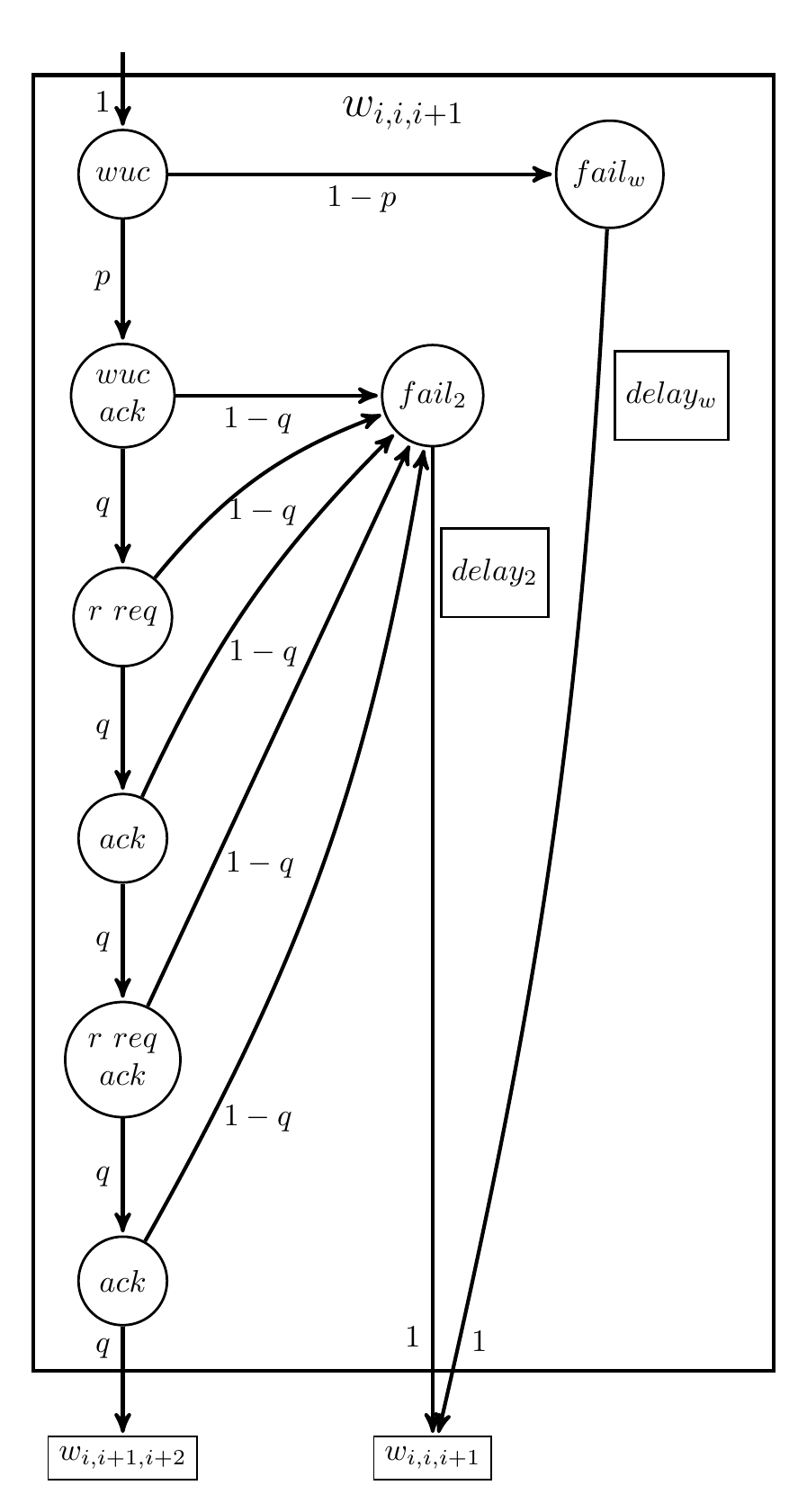}
		\caption{Markov chain for node $i$ (that also has the message to be delivered) attempting to wake-up node $i+1$}
		\label{fig:markov_wiii+1}
\end{figure}

The meta-models shown in Figures \ref{fig:t-rome_meta_model_1} to \ref{fig:t-rome_meta_model_6}, are composed of several Markov states as depicted in Figures \ref{fig:markov_tij} and \ref{fig:markov_wiii+1}. It can be seen in both Figures, that there exists a certain probability of success, but the attempts can also fail. In that case, a node enters a fail state that is exited with probability $1$ but has a certain delay connected to it. The delay just equals the timeout of the radio which is little more than the time required for the success case.

Here, we only show the Markov chains for the cases $T_{i,j}$ and $w_{i,j,j+1}$, as the chains for the cases $T_{i,m}$ and $w_{i,i,i+1}$, $w_{m-1,m-1,m}$ and $w_{i,m-1,m}$ are similar and can be achieved by plugging the Markov model into the corresponding meta model shown above. 

\subsubsection{Analysis}
\label{sec:analysis_trome}
Now, we can analyze the Markov Models with respect to the expected required time to send a messages via $m$ nodes. From Figures \ref{fig:markov_tij} and \ref{fig:markov_wiii+1} we can extract the expected times for all $T_{i,j}$ and $w_{i,j,j+1}$ states. $E[T_{i,j}]$ can be expressed by Equation (\ref{eq:e_time_tx}):
\begin{equation}
\label{eq:e_time_tx}
\begin{aligned}
E[T_{i,j}] = q^2 (T_{tx1} + E[w_{j,j,j+1}]) \\+ q(1-q)(T_{tx2} + E[w_{i,i,i+1}]) \\+ (1-q)(T_{tx3} + E[w_{i,i,i+1}]).
\end{aligned}
\end{equation}
Here, $T_{tx1}$ to $T_{tx3}$ are the times required to send the required communication packets in case of success ($T_{tx1}$), or the delay times required in case of failure ($T_{tx2}$ and $T_{tx3}$). $E[w_{j,j,j+1}]$ and $E[w_{i,i,i+1}]$ depict the expected times required in the corresponding T-ROME meta states which are given in Equations (\ref{eq:e_time_w1}) to (\ref{eq:e_time_w2}), below and can be extracted from Figure \ref{fig:markov_wiii+1}:
\begin{equation}
\label{eq:e_time_w1}
\begin{aligned}
E[w_{i,j,j+1}] = pq^5 (T_{w1} + E[w_{j,j+1,j+2}]) \\+ p(1-q^5)(T_{w2} + E[T_{i,j}]) \\+ (1-p)(T_{w3} + E[T_{i,j}]).
\end{aligned}
\end{equation}
Here, $T_{w1}$ to $T_{w3}$ are the times required to send the required communication packets in case of success ($T_{w1}$), or the delay times required in case of failure ($T_{w2}$ and $T_{w3}$). As we are looking on the general case of node $j$ attempting to wake-up node $j+1$ and the message is still at node $i$, the message will be send to node $j$ in case of failure (Figure \ref{fig:t-rome_meta_model_2}). Looking at the case where node $i$ has the message and attempts to wake-up node $i+1$, we find Equation (\ref{eq:e_time_w2}):
\begin{equation}
\label{eq:e_time_w2}
\begin{aligned}
E[w_{i,i,i+1}] = pq^5 (T_{w1} + E[T_{i,i+1}]) \\+ p(1-q^5)(T_{w2} + E[w_{i,i,i+1}]) \\+ (1-p)(T_{w3} + E[w_{i,i,i+1}]).
\end{aligned}
\end{equation}
Finally, we need to consider the cases where node $m-1$ attempts to wake-up node $m$, and the case where node $i$ attempts to send data to node $m$. These two cases are given by Equations (\ref{eq:e_time_w3}) and (\ref{eq:e_time_tx2}) for the wake-up and communication cases, respectively:
\begin{equation}
\label{eq:e_time_w3}
\begin{aligned}
E[w_{m-1,m-1,m}] = pq^5 (T_{w1} + E[T_{m-1,m}]) \\+ p(1-q^5)(T_{w2} + E[w_{m-1,m-1,m}]) \\+ (1-p)(T_{w3} + E[w_{m-1,m-1,m}])
\end{aligned}
\end{equation}
and 
\begin{equation}
\label{eq:e_time_tx2}
\begin{aligned}
E[T_{i,m}] = q^2 (T_{tx1}) + q(1-q)(T_{tx2} + E[w_{i,i,i+1}]) \\+ (1-q)(T_{tx3} + E[w_{i,i,i+1}]).
\end{aligned}
\end{equation}
Equations (\ref{eq:e_time_tx}) to (\ref{eq:e_time_tx2}) are a set of linear equations, which can be solved for certain $T_{txi}$ and $T_{wi}$ for $i=1,2,3$.

\subsection{Naive Algorithm}
\begin{figure}[ht!]
\centering
		\includegraphics[scale = 0.8]{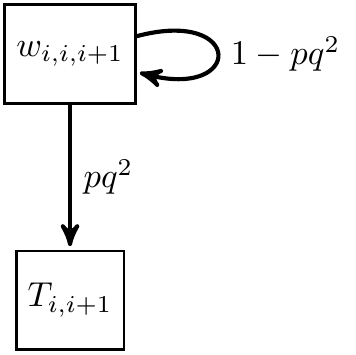}
		\caption{Meta model for node $i$ attempting to wake-up to node $i+1$ using the naive algorithm.}
		\label{fig:markov_singlehop_1}
\end{figure}

\begin{figure}[ht!]
\centering
		\includegraphics[scale = 0.8]{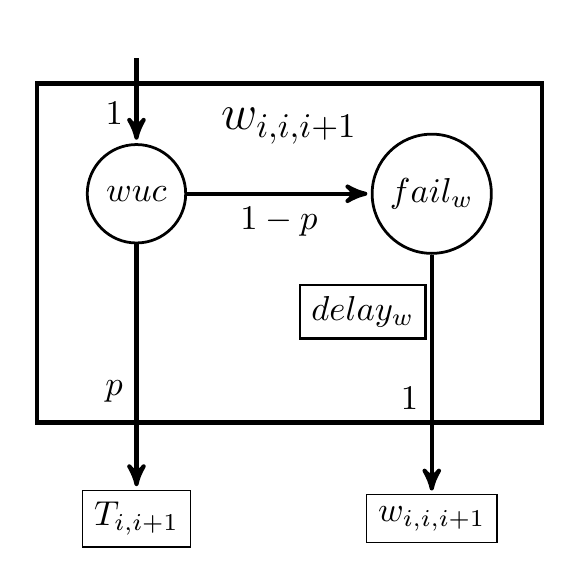}
		\caption{Markov chain for node $i$ (that also has the message to be delivered) attempting to wake-up node $i+1$ using the naive algorithm.}
		\label{fig:single_markov_wiii+1}
\end{figure}

\begin{figure}[ht!]
\centering
		\includegraphics[scale = 0.8]{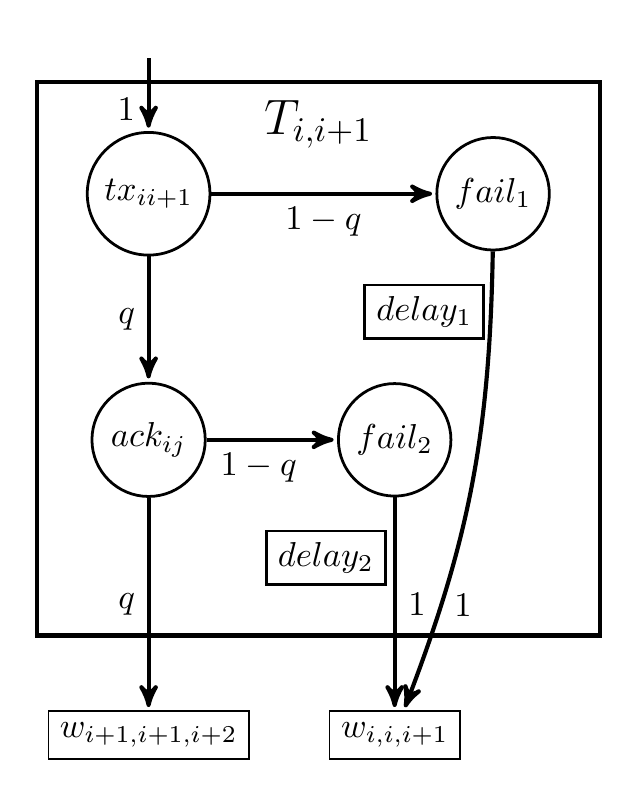}
		\caption{Markov chain for node $i$ attempting to send data to node $i+1$ using the naive algorithm}
		\label{fig:single_markov_t_ij}
\end{figure}

The naive algorithm wakes up and transmits data from node to node. Here, we assume following communication scheme: node $i$ sends a wake-up call to node $i+1$ directly followed by the data packet. Node $i+1$ acknowledges the data packet if it was received successfully. Figures \ref{fig:markov_singlehop_1} to \ref{fig:single_markov_t_ij} show the corresponding Markov models. 

Analysis of the models can be done similar to the analysis of T-ROME shown above in Section \ref{sec:analysis_trome}.

\subsection{CTP-WUR}
CTP-WUR protocol for unicast type packets is performed as introduced in \cite{basagni2016ctp} and presented in Section \ref{sec:protocols}. As it is possible to relay a message along one node, the CTP-WUR algorithm mainly consists of the meta models shown in Figures \ref{fig:t-rome_meta_model_1} (for single-hop transmissions) and \ref{fig:t-rome_meta_model_2} (for relaying, and: $j=i+1$). The Markov chain for wake-up is similar to the one shown in Figure \ref{fig:markov_wiii+1} but it is followed either by transmission ($T_{i,i+1}$) in case of single-hop, or by the next wake-up ($w_{i,i+1,i+2}$) in case of relaying. As the states are very similar to the already introduced models, we do not present them here in detail. Also, the system of linear equations is similar to that represented by Equations  (\ref{eq:e_time_tx}) to (\ref{eq:e_time_tx2}) above, and can be obtained easily by using the same techniques.

\section{Results}
\label{sec:results}
\subsection{Model Verification}
\begin{figure}[ht!]
\centering
		\includegraphics[scale=5]{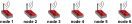}
		\caption{Deployment map of the sensor nodes during model verification.}
		\label{fig:verification_setup}
\end{figure}

To verify the models introduced in Section \ref{sec:analysis}, we conducted several experiments consisting of two, three, four, five and six nodes as shown in Figure \ref{fig:verification_setup} that shows the deployment of the sensor nodes during the test setups. As all tests were performed indoors, the nodes were placed at a distance of 1 m to each other and the output power of the main radio was adjusted to - 6 dBm and for the wake-up radio to 0 dBm. 

\begin{figure}[ht!]
\centering
		\includegraphics[scale=0.17]{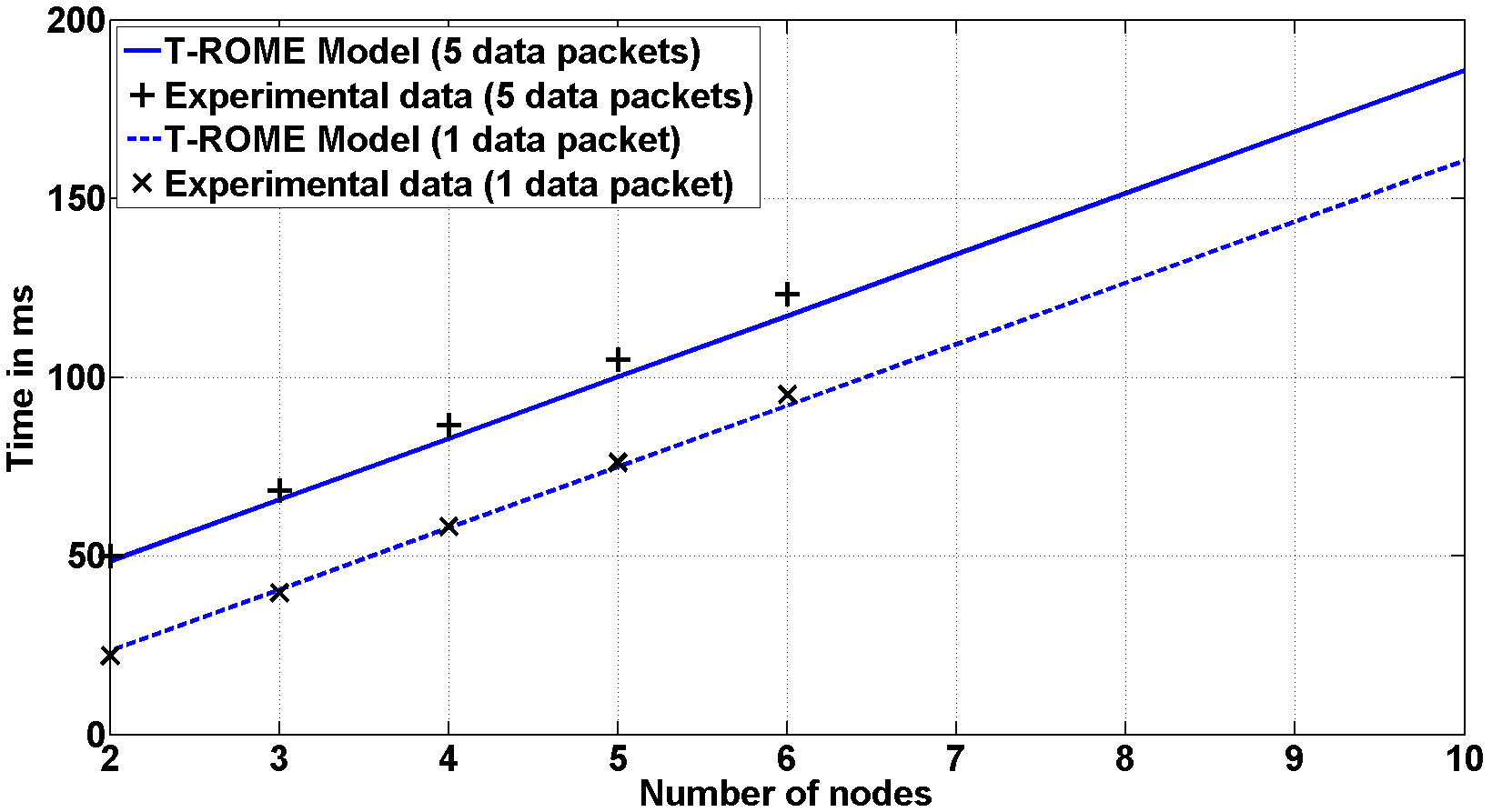}
		\caption{Simulated time required to send 1 data packet (dashed) and 5 data packets (solid) along several nodes assuming $p=q=1$. The points are data taken from the test setup}
		\label{fig:cmp_trome_exp}
\end{figure}

Node $1$ was always the source node and nodes $2$ to $6$ were either relay or sink nodes, depending on the experiment. In the experiment, each node could wake-up only its direct predecessor, that is, node $1$ could wake-up node $2$, node $2$ was able to wake-up node $3$ and so on. With respect to data communication, each node could communicate to each other (if awake). In the first experiment, node $1$ (source) had one message consisting of 100 byte to deliver to the sink (node $2$ to node $6$). In the second experiment, the sink had 5 messages each consisting of 100 bytes to deliver. In each experiment, we were running the T-ROME algorithm and measured the required time until the message was delivered at the sink. Figure \ref{fig:cmp_trome_exp} shows the experimental data for 1 message as crosses and for 5 data packets as pluses. The curves in Figure \ref{fig:cmp_trome_exp} show the expected times by using the T-ROME Markov models of Section \ref{sec:analysis} above, assuming $p=q=1$. Figure \ref{fig:cmp_trome_exp} show that both, the expected and the measured times correspond very well.

\subsection{Performance Analysis}
\begin{figure}[ht!]
\centering
		\includegraphics[scale=0.17]{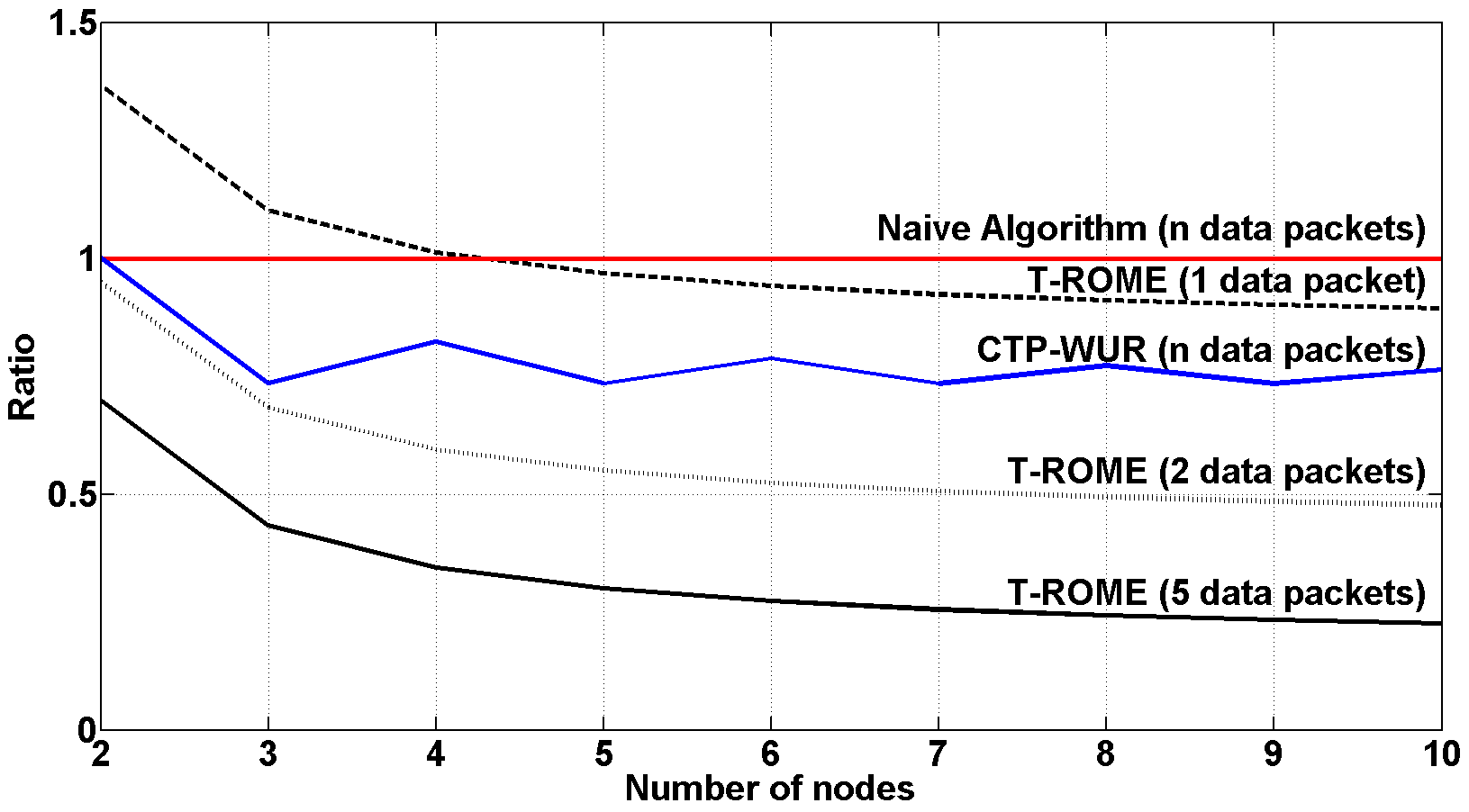}
		\caption{Simulated time performance of T-ROME and CTP-WUR compared to the naive algorithm, assuming $p=q=1$. Black: T-ROME for 1 (dotted), 2 (dashed) and 5 data packets (solid), red: naive algorithm and blue: CTP-WUR.}
		\label{fig:cmp_time_1}
\end{figure}

After model verification, we compared T-ROME, CTP-WUR and the naive algorithm using the models introduced in Section \ref{sec:analysis}. Figure \ref{fig:cmp_time_1} shows the simulation results with respect to the required time to send 1, 2 and 5 data packets along several nodes, assuming $p=q=1$. All results are compared to the performance of the naive algorithm. It can be seen that CTP-WUR performs equally good as the naive algorithm in case of two participating nodes and performs about a constant ratio better than the naive algorithm for more than two participating nodes. This behavior is expected since CTP-WUR uses the same communication messages as the naive algorithm and saves a constant amount of time by using one relay node. For both algorithms (naive and CTP-WUR), the ratio stays constant, regardless of how many data packets are sent.

Looking at the results for sending one data packet with T-ROME, we find that it requires more time than the naive algorithm (factor of around 1.4) for two participating nodes due to the additional messages required in the protocol. For four participating nodes, T-ROME performs equally good as the naive algorithm and for more than four nodes, it outperforms it but it does not reach the performance of CTP-WUR. However, T-ROME outperforms CTP-WUR and the naive algorithm, when delivering two or more data packets with two or more participating nodes due to the savings of relaying. 

\begin{figure}[ht!]
\centering
		\includegraphics[scale=0.17]{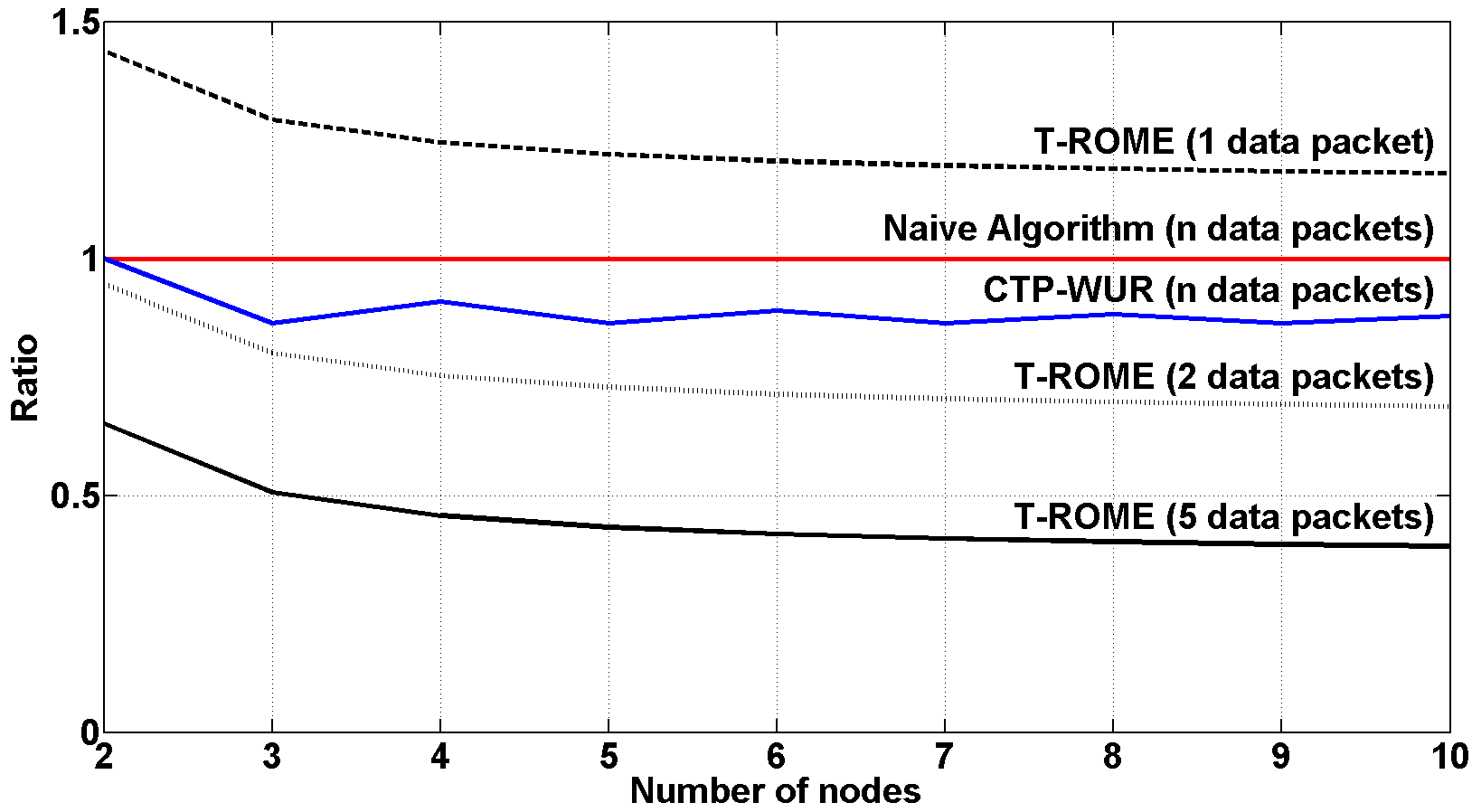}
		\caption{Simulated time performance of T-ROME and CTP-WUR compared to the naive algorithm, assuming $p=0.75$ and $q=0.97$. Black: T-ROME for 1 (dotted), 2 (dashed) and 5 data packets (solid), red: naive algorithm and blue: CTP-WUR.}
		\label{fig:cmp_time_pq}
\end{figure}

Figure \ref{fig:cmp_time_pq} shows the time performance comparison for $p=0.75$ and $q=0.97$. It can be seen that the differences between the naive and the other (T-ROME and CTP-WUR) are getting smaller and that T-ROME outperforms CTP-WUR and the naive algorithm only when more than one data packet is sent. The sensitivity of T-ROME originates from the larger amount of communication packets as required by the protocol.

Additional issues to consider when analyzing the performance of T-ROME are opportunistic routing approaches and route adjustments based on link quality estimation that are used in some of the wake-up protocols introduced in Section \ref{sec:protocols}. As these parameters are not yet implemented in T-ROME, the performance of T-ROME could decrease due to an increased number of packet retransmissions. We expect T-ROME to gain a similar or even bigger advantage from these features, compared to other protocols as T-ROME provides additional flexibility during candidate coordination.

\subsection{Energy Budget}

\begin{figure}[ht!]
\centering
		\includegraphics[scale=0.7]{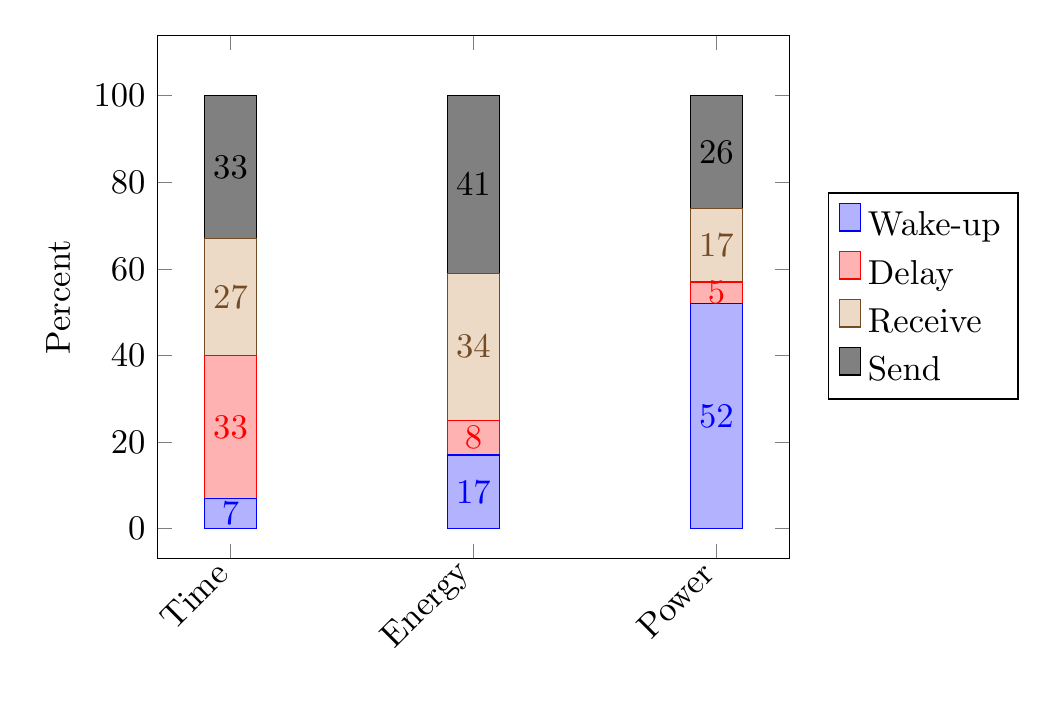}
		\caption{Time, energy and power requirements of the sender node. }
		\label{fig:source_sending}
\end{figure}

Figure \ref{fig:source_sending} shows the time, energy and power shares of the source node (node 13) of the proposed protocol calculated with the numbers provided above (TTL = 3, 5 data packets each 100 byte large). It can be seen that the wake-up call requires only 7 \% of the time and delay, receive and send share around a third of the time. Looking at the energy shares of each state, it is evident that sending and receiving require most of the energy but the wake-up call still needs almost a fifth of the total energy. Power consumption during wait periods is less than 10 \% of the total amount. But due to the high current required during sending wake-up messages, more than 50 \% of the allocated power is consumed.

Table \ref{tab:energy_consumption} shows the energy consumption of each node using the proposed protocol for a network consisting of four nodes and TTL=3. It can be seen that the source node requires most energy and the sink node fewest, as it does not send a wake-up message and has no delay states. Here, the delay is the time required by the sender to give the forwarding nodes time to wake-up their neighbors. So the delay consists of the times required to send the packets WUC, WUC ACK, R\_REQ, and ACK.

\begin{table}[h!]
  \caption{Energy consumption of a network of four nodes (TTL = 3) in mJ of each node for states sending WUC (wake-up call), delay, receive data and transmit data at 3.3 V, when transmitting 500 byte in 5 data packets (100 byte per packet) directly from source to sink.}
\label{tab:energy_consumption}
  \begin{tabularx}{\hsize}{p{1.65cm} X X X X X}
\toprule
    node ID & WUC [mJ]& Delay [mJ]& Receive [mJ]& Send [mJ] & total [mJ]\\
\midrule
    13 (source) & 0.6 & 0.3 & 0.8 & 1.6  & 3.3 \\
    12 (relay 1) &0.6 & 0.1 & 0.7 & 0.3 & 1.7  \\
    11 (relay 2)&0.6 & - & 0.7 & 0.3 & 1.6  \\
    10 (sink) & - & - & 1.5 & 0.6 & 2.1\\
\bottomrule
  \end{tabularx}
\end{table}

\begin{figure}[ht!]
\centering
		\includegraphics[scale=0.17]{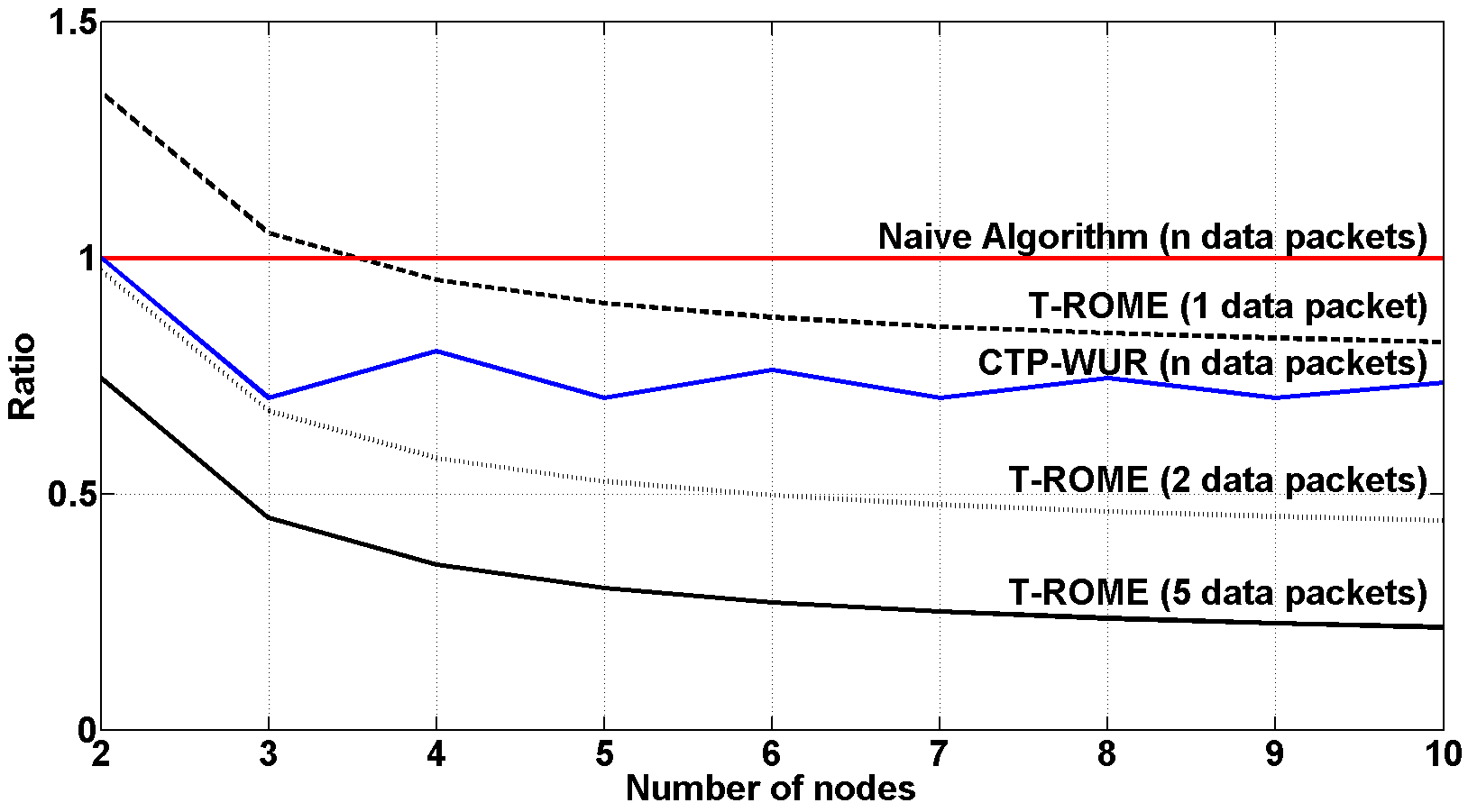}
		\caption{Simulated energy performance of T-ROME and CTP-WUR compared to the naive algorithm, assuming $p=q=1$. Black: T-ROME for 1 (dotted), 2 (dashed) and 5 data packets (solid), red: naive algorithm and blue: CTP-WUR.}
		\label{fig:cmp_energy_1}
\end{figure}

Figure \ref{fig:cmp_energy_1} shows the simulated energy performance analysis of CTP-WUR and T-ROME compared to the performance of the naive algorithm, for $p=q=1$. It can be seen that the energy performance is similar to the time performance. CTP-WUR outperforms the naive algorithm for a communication with more than two nodes. Due to relaying, T-ROME outperforms the naive protocol and CTP-WUR as soon as more than one data packet is sent.  

\subsection{Overhead}
\begin{figure}[ht!]
\centering
		\includegraphics[scale=0.17]{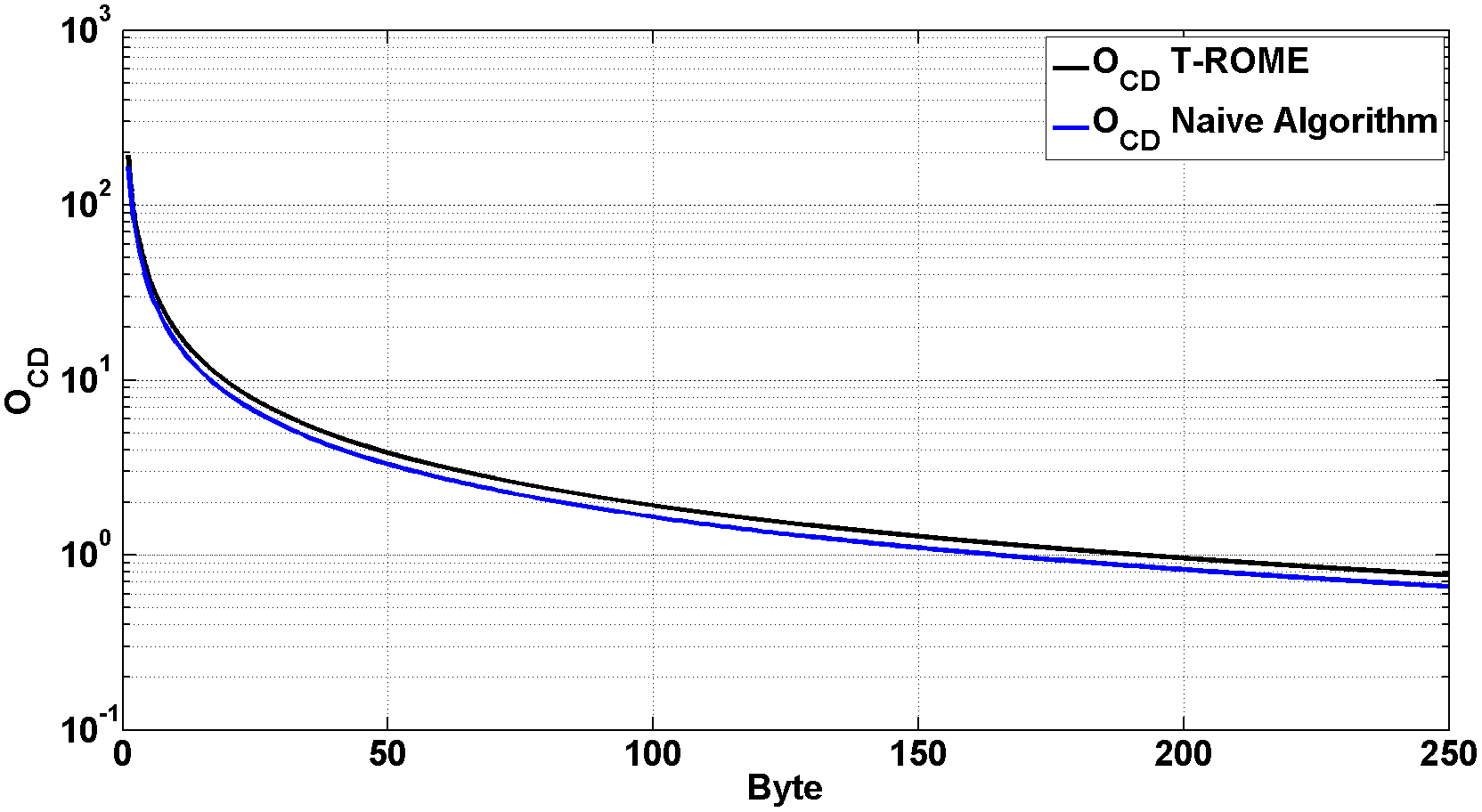}
		\caption{T-ROME overhead (blue) compared to the overhead of the naive algorithm (black) for sending one data packet}
		\label{fig:ocd_one}
\end{figure}

By using the control overhead ratio $O_{CD}$ as the ratio of control bit sent ($CB_S$) over data bit delivered ($DB_D$) ($OC_D = CB_S/DB_D$), we analyzed T-ROME and the naive algorithm with respect to protocol overhead. Figure \ref{fig:ocd_one} shows $O_{CD}$ over data for transmission of one data packet. It can be seen that transmission of few bytes requires a large overhead due to a large amount of byte required for the wake-up call (162 bytes). It can also be seen that the naive algorithm requires less overhead than T-ROME due to a fewer number of control packets. The naive algorithm reaches the brake-even point at 165 bytes and T-ROME at 192 bytes. 

\begin{figure}[ht!]
\centering
		\includegraphics[scale=0.17]{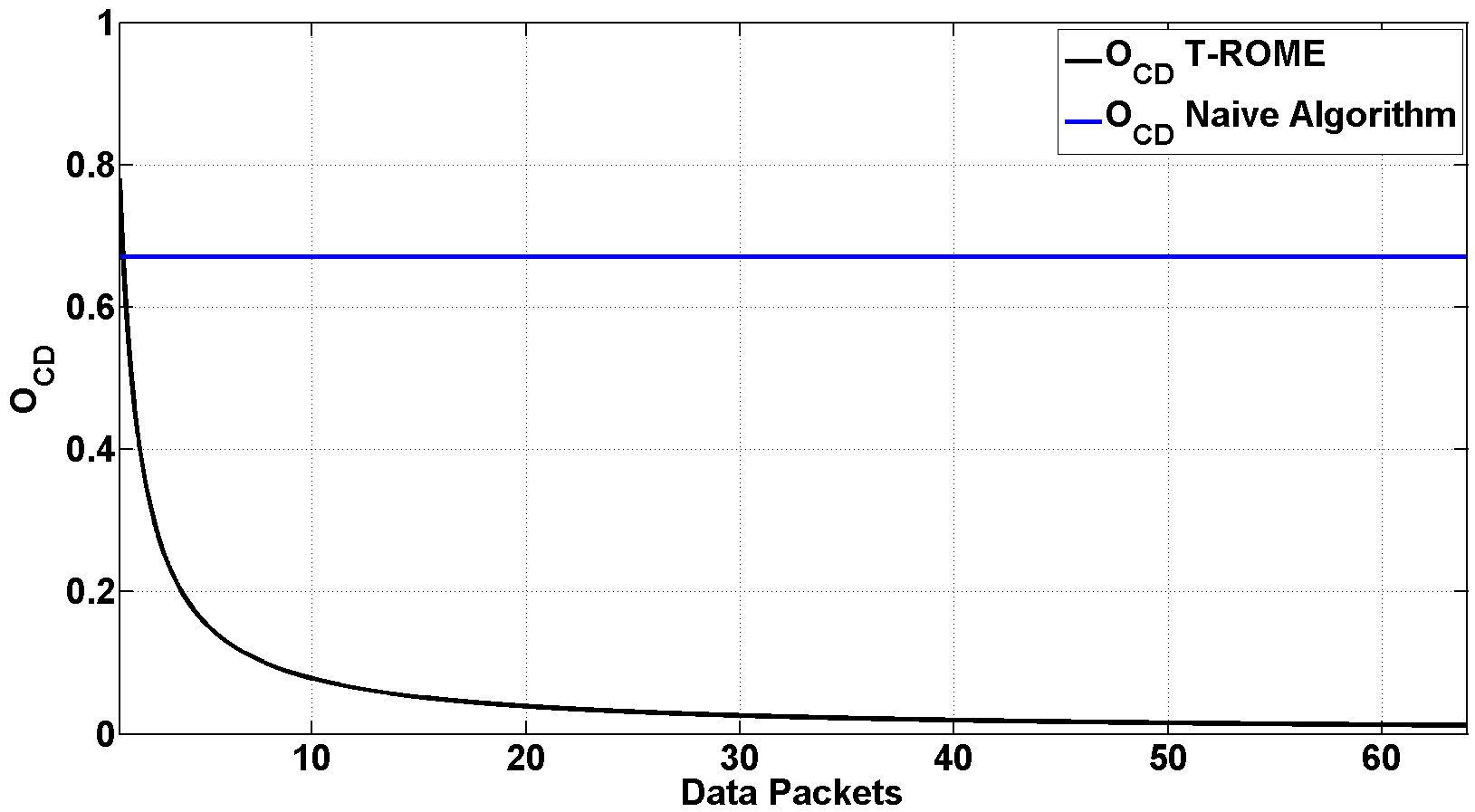}
		\caption{T-ROME overhead (blue) compared to the overhead of the naive algorithm (black) for sending 64 data packets}
		\label{fig:ocd_64}
\end{figure}

Figure \ref{fig:ocd_64} shows $O_{CD}$ plotted against number of sent packets. As T-ROME requires almost no further control byte after a link is established, $O_{DC}$ decreases quickly for sending of more than one data packet. In the case of the naive algorithm, $O_{CD}$ stays constant as every packet requires the same amount of control byte.  

\section{Conclusions and Outlook}
\label{sec:outlook}

%
In this article, we developed for the first time an energy efficient and simple cross-layer network protocol for wireless wake-up sensor networks (T-ROME) followed by its comprehensive modeling and analysis. The protocol combines the advantages of wake-up receivers such as low-power consumption and on-demand communication together with the advantage of long-range communication radios, that is their superior sensitivity. 

T-ROME makes use of the different communication ranges of communication and wake-up radio. The protocol saves energy by skipping nodes during data communication. Furthermore, T-ROME introduces a set of parameters to optimize the relaying process by dynamically choosing the most appropriate stopover nodes in case the sink is not reachable within one communication hop. The total number of wake-up packets can be reduced with T-ROME by accumulating sensor data and sending up to 64 data packets (16 kbyte) in a row once a communication link is established.

Based on newly developed Markov chain models we analyzed T-ROME and other state-of-the-art communication protocols for wake-up receivers regarding energy consumption, communication duration and overhead. Especially, we investigated, modeled and analyzed CTP-WUR and a naive communication algorithm and compared their performance to that of T-ROME. We demonstrated that our proposed protocol outperforms existing protocols in many cases particularly by sending several data packets at once and by skipping nodes during communication.

Next steps will include the realization of a dynamic routing protocol based on the proposed protocol scheme as presented here. In addition, we will introduce further parameters that support the decision finding at the sender node, like link quality, receiver signal strength or remaining energy level at the receiver side to increase network stability and to enhance its lifetime.

\section*{Acknowledgments}

We gratefully acknowledge support from BASt (FE 88.0126/2012) who funded this research.

\section*{References}

\bibliography{protocols}

\end{document}